\documentclass[preprint,amsmath,amssymb,aps,prd,nofootinbib]{revtex4}

\usepackage{epsfig}
\begin{document}
\draft
\title{\mbox{}\\[10pt]
 Bridges of Low Energy observables with Leptogenesis \\
 in $\mu-\tau$ Reflection Symmetry}

\author{Y. H. Ahn$^{a,}$\footnote{yhahn@phys.sinica.edu.tw,~ skkang@snut.ac.kr,~ cskim@yonsei.ac.kr,~ phong@cskim.yonsei.ac.kr},~~
        Sin Kyu Kang$^{b,}$,~~ 
       C. S. Kim$^{c,}$,~~ 
        T. Phong Nguyen$^{c,}$
        }

\address{$^a$ Institute of Physics, Academia Sinica, Taipei, Taiwan 115, ROC\\
         $^b$ School of Liberal Arts, Seoul National Univ. of Technology, Seoul 139-743, Korea\\
         $^c$ Dept. of Physics and IPAP, Yonsei University, Seoul 120-749, Korea}




\begin{abstract}
\begin{small}
\noindent We consider an exact $\mu - \tau$  reflection symmetry in neutrino sector
realized at the GUT scale in the context of the seesaw model with and without supersymmetry.
Assuming the two lighter heavy Majorana neutrinos are degenerate at the GUT scale,
it is shown that the renormalization group (RG) evolution
from the GUT scale to the seesaw scale gives rise to breaking of the $\mu-\tau$ symmetry and
a tiny splitting between two degenerate heavy Majorana neutrino masses
as well as small variations of the CP phases in $Y_{\nu}$, which are essential to achieve a successful leptogenesis.
Such small RG effects lead to tiny deviations of $\theta_{23}$ from the maximal value  and the CP phase $\delta_{\rm CP}$
from $\frac{\pi}{2}$ imposed at the GUT scale due to $\mu-\tau$ reflection symmetry.
In our scenario, the required amount of the baryon asymmetry $\eta_B$ can be generated via so-called resonant $e$-leptogenesis,
in which the wash-out factor concerned with electron flavor plays a crucial role in reproducing a successful leptogenesis.
We show that CP violation responsible for the generation of baryon asymmetry of our universe
can be directly linked with  CP violation measurable through neutrino oscillation as well as neutrino
mixing angles $\theta_{12}$ and $\theta_{13}$. We expect that, in addition to the reactor and long baseline neutrino experiments,
the measurements for the supersymmetric parameter
$\tan\beta$ at future collider experiments would serve as an indirect test of our scenario of  baryogenesis
based on  the $\mu-\tau$ reflection symmetry.\\

\end{small}
\end{abstract}

\maketitle
\section{Introduction}

The present neutrino experimental data \cite{atm,SK2002,SNO} exhibit
that the atmospheric neutrino deficit points toward a maximal mixing
between the tau and muon neutrinos, however, the solar neutrino
deficit favors a not-so-maximal mixing between the electron and muon
neutrinos, $i.e.$, $|U_{\mu3}|\simeq|U_{\tau3}|\simeq1/\sqrt{2}$, where
$U$ is the leptonic PMNS mixing matrix. In addition, although we do
not have yet any firm evidence for the neutrino oscillation arisen
from the 1st and 3rd generation flavor mixing, there is a bound on
the mixing element $U_{e3}$ from CHOOZ reactor experiment,
$|U_{e3}|<0.2$ \cite{chooz}. Although neutrinos have gradually
revealed their properties in various experiments since the historic
Super-Kamiokande confirmation of neutrino oscillations \cite{atm},
properties related to the leptonic CP violation are completely
unknown yet.
CP violations in the leptonic sector are also obligatory, if the matter and antimatter asymmetry of the Universe is made through
leptogenesis scenario \cite{review} in the seesaw models \cite{seesaw}.

Recently, the so-called $\mu$-$\tau$ reflection symmetry \cite{Harrison:2002et,Grimus} has been imposed in light neutrino mass matrices so that the maximal atmospheric mixing $\theta_{23}=\pi/4$ as well as the maximal value for the Dirac CP violating phase $\delta_{\rm CP}=\pi/2$  are predicted \cite{A4CP,yasue}.  
Under the $\mu$-$\tau$ reflection symmetry in the context of a seesaw model, it is supposed that
the right-handed neutrinos transform into the charge conjugates of themselves, $N_{i}\rightarrow N^{c}_{i}$ and the light left-handed
neutrinos into $\nu_e \rightarrow \xi_1 \nu_e^{c}, \nu_\mu \rightarrow \xi_2 \nu_\tau^{c}, \nu_\tau \rightarrow \xi_2 \nu_\mu^{c}$
\cite{Harrison:2002et,Grimus,Farzan} with the corresponding phases $\xi_i$.
The effective mass matrix derived from the seesaw mechanism is invariant under the $\mu-\tau$ reflection symmetry, too.
 In the light of the CP violation from the neutrino oscillations \cite{factory}, the $\mu-\tau$ reflection symmetry
  indicates that, for given values of mixing angles, the CP asymmetry $P(\nu_{\mu}\rightarrow\nu_{e})-P(\bar{\nu}_{\mu}\rightarrow\bar{\nu}_{e})$
  is maximal.

In this paper, we consider a seesaw model with the $\mu-\tau$ reflection symmetry.
In contrast with other models concerned with the $\mu-\tau$ reflection symmetry, we impose the exact symmetry
at the GUT scale and consider renormalization group (RG) evolution  \cite{RGE,Antusch:2005gp,RG1,Antusch:2002rr} on the neutrino Dirac Yukawa couplings,
heavy Majorana neutrino masses and CP phases by running from the GUT scale to a seesaw scale.
One of interesting issues concerned with neutrino models with the $\mu-\tau$ reflection symmetry is if baryogenesis
can be successfully realized through leptogenesis or not.
In fact, high energy cosmological CP violation responsible for baryogenesis vanishes as long as the $\mu-\tau$ reflection symmetry is kept
exact and flavor effects associated with the charged leptons are not taken into account \cite{Grimus}.
However, it has been shown that flavor effects can lead to a successful leptogenesis for $M_{1}<10^{13}$ GeV in the limit of the
$\mu-\tau$ reflection symmetry as long as
a partial lepton asymmetry associated with a lepton flavor does not vanish even if total lepton asymmetry is zero \cite{Farzan}.
According to our numerical estimates made in this paper, it is not true in some cases of  heavy Majorana neutrino mass spectrum with
hierarchy structures like $M_{1}\ll M_{2}<M_{3}$ and $M_{1}\simeq M_{2} << M_{3}$.
In such cases, as will be shown later, breaking of the $\mu-\tau$ reflection symmetry can lead to successful leptogenesis.
Instead of introducing {\it ad-hoc} soft symmetry breaking terms,
in this paper, we consider a possibility that RG evolutions of neutrino parameters can beak the $\mu-\tau$ reflection symmetry and then
examine if such  breaking effects due to  the RG evolutions can lead to successful baryogenesis.
To do that, we simply assume that two lighter heavy Majorana neutrino masses are exactly degenerate at the GUT scale.
If a mass splitting between two degenerate heavy Majorana neutrinos due to the RG evolution comes out to be small,
we expect that the lepton asymmetry can be resonantly enhanced \cite{Pilaftsis:1997jf}.
However, we show that even though the total lepton asymmetry can be resonantly enhanced,
the magnitude of the total lepton asymmetry is not enough to achieve successful baryogenesis
in the context of the Standard Model (SM) with the $\mu-\tau$ reflection symmetry broken by RG corrections.

In general, the absolute magnitudes of the lepton asymmetries for each lepton flavor can be larger than that of the total lepton asymmetry.
Furthermore, since the interactions mediated by the tau and muon coupling are in thermal equilibrium
at below the temperature $M\sim10^{9}$ GeV,  the processes which wash out lepton number are flavor dependent and thus
the lepton asymmetries for each flavor should be treated separately with different wash-out factors.
It has been known that flavor effects can enlarge the area of parameter space where leptogenesis can work.
Thus, it is meaningful to investigate if leptogenesis including flavor effects can be successfully realized
in the case that  leptogenesis without flavor effects does not work successfully.
In this paper, we shall examine if  the flavor effects on leptogenesis can help to achieve successful baryogenesis
in the context of the SM with the $\mu-\tau$ reflection symmetry broken by RG corrections.
In particular, we shall carefully discuss how the wash-out factors for each lepton flavor can be significant to achieve successful leptogenesis.

In the supersymmetric seesaw model (SSM), the baryon asymmetry of our universe can be achieved
leptogenesis via the decays of heavy Majorana neutrinos.
One of interesting points is that the lepton asymmetry in the SSM can be enhanced in the case of large $\tan\beta$
because it is proportional to the charged lepton Yukawa couplings $Y_{l}=Y_{l_{\rm SM}}(1+\tan\beta)$.
Thus, even for the parameter space where leptogenesis in the SM does not work, supersymmetry can make
leptogenesis working.
We also expect that since the RG evolutions in the SSM are different from those in the SM, the parameter space
where leptogenesis can work must be different each other.
In this paper, we shall study how the results concerned with leptogenesis in the SM can change when we supersymmetrize
it.

This paper is organized as follows:
In Sec. II, we present a seesaw model
reflecting $\mu-\tau$ reflecting symmetry  at the GUT scale in the context of both SM and SSM.
In Sec. III, we discuss how we can obtain a RG-improved neutrino mass matrix. We show that the RG evolutions break the $\mu-\tau$
reflection symmetry and derive the deviations of low energy neutrino mixing angles
and CP phase from their tree level values which can be parameterized in terms of a parameter concerned with the RG evolution.
The discussion for
RG evolution from the GUT scale to low scale and useful formulae are given in Appendix. In Sec.
IV, we show how successful leptogenesis can be radiatively induced in our scheme.
Here, we discuss how lepton flavor effects are important to achieve successful leptogenesis and carefully estimate
the wash-out factors by taking flavor effects into account.
 Numerical results and conclusion are given in Sec. V.

\section{Seesaw Model with the $\mu-\tau$ Reflection symmetry}

\subsection{In case of the SM}

To begin with, let us consider the Lagrangian of the lepton sector
from which the seesaw mechanism works,
\begin{eqnarray}
  {\cal L}_{m}=-\overline{L}\textbf{Y}_{\nu}N\tilde{\phi}
              - \overline{L}\textbf{Y}_{l} l_{R} {\phi}
-\frac{1}{2}\overline{N}^{c} \textbf{M}_{R} N+h.c.,
\label{lagrangian}
\end{eqnarray}
where the family indices have been omitted and $L_i,~ l_R,~ \phi,~ N$,
$i=e,\mu, \tau$ stand for SU(2) lepton doublet fields, charged lepton singlet fields and
Higgs scalar and singlet heavy Majorana neutrino, respectively.
 In the above lagrangian,
$\textbf{M}_R$, $\textbf{Y}_l$ and $\textbf{Y}_{\nu}$ are the $3\times 3$ heavy Majorana neutrino mass matrix, charged
lepton and neutrino Dirac Yukawa matrices, respectively. After spontaneous electroweak symmetry breaking, the seesaw mechanism leads to the
following effective light neutrino mass term,
\begin{eqnarray}
  &&m_{\rm eff}=-\textbf{Y}^{T}_{\nu}\textbf{M}^{-1}_{R}\textbf{Y}_{\nu}\upsilon^{2} ~,
  \label{meff}
\end{eqnarray}
where $\upsilon$ is a vacuum expectation value of the Higgs
field with $\upsilon\approx174$ GeV.

Let us impose the $\mu-\tau$ reflection symmetry for the neutrino sectors at the GUT scale,
which is the combined operation of $\mu-\tau$ flavor exchange in PMNS mixing matrix and CP transformation on the
leptonic sector \cite{Harrison:2002et}, in the basis where both
the charged lepton mass and heavy Majorana mass matrices are
diagonal.
Under the $\mu-\tau$ reflection symmetry,
the right-handed neutrinos transform into the charge conjugates of
themselves, $N_{i}\rightarrow N^{c}_{i}$ and thus the
elements of $\textbf{M}_R$ should be real.
Then, the neutrino
Dirac Yukawa matrix and the heavy Majorana neutrino mass matrix, at the GUT scale,
are taken to be
 \begin{eqnarray}
  \textbf{Y}_{\nu}= {\left(\begin{array}{ccc}
 a_{1} &  b_{1}e^{i\phi_{1}} &  b_{1}e^{-i\phi_{1}} \\
 a_{2} &  b_{2}e^{i\phi_{2}} &  b_{2}e^{-i\phi_{2}} \\
 a_{3} &  b_{3}e^{i\phi_{3}} &  b_{3}e^{-i\phi_{3}}
 \end{array}\right)}~,
 ~~~~~\textbf{M}_{R}={\left(\begin{array}{ccc}
 M_{1} & 0 &  0 \\
 0 & M_{2} & 0 \\
 0 & 0 & M_{3} \end{array}\right)}~,
 \label{input1}
 \end{eqnarray}
where $a_i, b_i ~(i = 1, 2, 3) $ of the neutrino Dirac Yukawa matrix are
all real. We assume that the two heavy Majorana neutrinos are exactly degenerate in mass, which
is much smaller than that of the right-handed neutrino $N_3$, {\it i.e.} $M_{1}=M_{2}\equiv M\ll M_{3}$.
It turns out, after our analysis, that in the light of seesaw mechanism with such a hierarchy of heavy Majorana neutrinos
the effects of the parameters $a_3$ and $\phi_3$ on low energy neutrino phenomenology and even leptogenesis are negligibly small.
Thus, for our convenience, we take $a_3 =0$ and $\phi_{3}=0$ at the GUT scale.

Introducing several parameters defined by the ratios among the parameters appeared in Eq.~(\ref{input1}), as follows,
  \begin{eqnarray}
  m_{0}\equiv\upsilon^{2}\frac{b^{2}_{3}}{M}~,~~~\lambda\equiv\frac{a_{1}}{b_{3}}~,~~~\omega\equiv\frac{b_{1}}{b_{3}}~,
  ~~~\chi\equiv\frac{a_{2}}{b_{3}}~,~~~\kappa\equiv\frac{b_{2}}{b_{3}}~,~~~\eta\equiv\frac{M_{3}}{M}~,
  \label{input2}
  \end{eqnarray}
the effective light neutrino mass matrix generated through
seesaw mechanism can be written
as
 \begin{eqnarray}
  m_{\rm eff} = m_{0}{\left(\begin{array}{ccc}
  m_{ee} &   m_{e\mu} &  m^{\ast}_{e\mu} \\
  m_{e\mu} &  m_{\mu\mu} &  m_{\mu\tau} \\
  m^{\ast}_{e\mu} &  m_{\mu\tau} &  m^{\ast}_{\mu\mu} \end{array}\right)}~,
 \label{meff1}
\end{eqnarray}
where
 \begin{eqnarray}
  m_{ee} &=& \lambda^{2}+\chi^{2}~,~~~~~~~~~~~~~~~~~~~~~~~~~
  m_{e\mu} = \lambda\omega e^{i\phi_{1}}+\chi\kappa e^{i\phi_{2}}\nonumber\\
  m_{\mu\mu} &=& \omega^{2}e^{2i\phi_{1}}+\kappa^{2}e^{2i\phi_{2}}+\frac{1}{\eta}~,~~~~~~
  m_{\mu\tau} = \omega^{2}+\kappa^{2}+\frac{1}{\eta}~.\nonumber
\end{eqnarray}
We see that the effective neutrino mass matrix $m_{\rm eff}$ reflects the $\mu-\tau$
reflection symmetry \cite{Harrison:2002et}.
Here, we have not yet considered RG corrections to $m_{\rm eff}$ from the GUT scale to a seesaw scale, so
the form of $m_{\rm eff}$ given in Eq.~(\ref{meff1}) is regarded at tree level.

It is not difficult to see that the seesaw model based on
Eq.~(\ref{input1}) leads to  the normal hierarchical light
neutrino mass spectrum because we take diagonal form of heavy
Majorana neutrino mass matrix \cite{Ahns1,Ahn:2006rn,SFKing}.
Thus, the RG effects from a seesaw scale to electroweak scale on $m_{\rm eff}$ as well as on the neutrino mixing matrix
$U_{\rm PMNS}$ are expected to be very small.
Starting from the $\mu-\tau$ reflection symmetric forms of $Y_{\nu}$ and $M_R$, we shall show that the $\mu-\tau$ reflection symmetry
is broken due to the RG running effects between the GUT and a seesaw scale and they can lead to
successful leptogenesis without being in conflict with experimental low energy constraints.

Concerned with CP violation, we notice from Eq.~(\ref{meff1}) that the CP phases $\phi_{1,2}$ coming from $\textbf{Y}_{\nu}$
take part in low-energy CP violation.
To see how the CP phases $\phi_{1,2}$ are correlated with low energy CP violation measurable through neutrino oscillations,
let us consider CP violation parameter defined through Jarlskog invariant \cite{Jarlskog}
\begin{eqnarray}
    J_{\rm CP} &=& \frac{1}{8}\sin2\theta_{12}\sin2\theta_{23}\sin2\theta_{13}\cos\theta_{13}\sin\delta_{\rm CP}
    = \frac{{\rm Im}[h_{e\mu}h_{\mu\tau}h_{\tau e}]}{\Delta m^{2}_{21}\Delta m^{2}_{31}\Delta m^{2}_{32}},
 \label{CP0}
 \end{eqnarray}
 where $h=m^\dagger_{\rm eff} m_{\rm eff}$ and $\Delta m^{2}_{ij} = m^2_i - m^2_j$.
{}From the matrix given in Eq.~(\ref{meff1}) it can be written  as
 \begin{eqnarray}
  J_{\rm CP}\sim{\rm Im}[h_{e\mu}h_{\mu\tau}h_{\tau e}]\simeq m^{6}_{0}A\kappa\omega\sin\Delta\phi_{12}+O(\eta^{-1}),
 \end{eqnarray}
where $A$ is the lengthy function composed of the parameters $\kappa,\omega,\chi,\lambda,\cos\Delta\phi_{12}$
and $\Delta \phi_{ij} = \phi_i-\phi_j$.
We see from the above equation that $J_{\rm CP}$ is correlated with high energy CP parameter $\Delta \phi_{12}$.
As long as $\Delta\phi_{12}\neq0$ and $A\neq0$,  $J_{\rm CP}$ have non-vanishing value, which would be a signal of CP violation.

The combination of Dirac neutrino Yukawa matrices, which is relevant for leptogenesis, is given by
 \begin{eqnarray}
  \textbf{Y}_{\nu}\textbf{Y}^{\dag}_{\nu}= H &=& b^2_3\left(\begin{array}{ccc}
  \lambda^2+2\omega^2   &  \lambda\chi+2\omega\kappa\cos\Delta\phi_{12}  & 2\omega\cos\Delta\phi_{13}  \\
  \lambda\chi+2\omega\kappa\cos\Delta\phi_{12} & \chi^2+2\kappa^2 & 2\kappa\cos\Delta\phi_{23} \\
  2\omega\cos\Delta\phi_{13}  &  2\kappa\cos\Delta\phi_{23} & 2 \\
  \end{array} \right)~.
  \label{H1}
 \end{eqnarray}
{}From this, we find that the hermitian quantity $\textbf{Y}_{\nu}\textbf{Y}^{\dag}_{\nu}$ in the limit of the $\mu-\tau$ reflection
symmetry leads to ${\rm Im}[\textbf{Y}_{\nu}\textbf{Y}^{\dag}_{\nu}]=0$ and thus vanishing lepton asymmetry which is undesirable
for a successful leptogenesis.
To generate non-vanishing lepton asymmetry, both the mass degeneracy of the 1st and 2nd heavy Majorana neutrinos and the $\mu-\tau$
reflection symmetric texture of $\textbf{Y}_{\nu}$ in Eq.~(\ref{input1}) should be broken.

\subsection{In case of the SSM}

In the SSM,  the leptonic superpotential is given  by
\begin{eqnarray}
  W_{\rm lepton}=\hat{l}^{c}_{L}\textbf{Y}_{l}\hat{L}\cdot \hat{H}_{d}
              +\hat{N}^{c}_{L}\textbf{Y}_{\nu} \hat{L}\cdot \hat{H}_{u}
-\frac{1}{2}\hat{N}^{cT}_{L} \textbf{M}_{R} \hat{N}^{c}_{L}~,
\label{lagrangian2}
\end{eqnarray}
where the family indices have been omitted and $\hat{L}$ stands for the chiral super-multiplets of the ${\rm SU(2)}_{L}$ doublet
lepton fields, $\hat{H}_{u,d}$ are the Higgs doublet fields with hypercharge $\pm1/2$, $\hat{N}^{c}_{L}$ and $\hat{l}^{c}_{L}$ are the super-multiplet of the ${\rm SU(2)}_{L}$ singlet neutrino and charged lepton field, respectively.
 After spontaneous electroweak symmetry breaking, the seesaw mechanism leads to a
following effective light neutrino mass term,
\begin{eqnarray}
  &&m_{\rm eff}=\textbf{Y}^{T}_{\nu}\textbf{M}^{-1}_{R}\textbf{Y}_{\nu}\langle H_{u}\rangle^{2} ~,
  \label{meff2}
\end{eqnarray}
where $\langle H_{u}\rangle^{2}=\upsilon^{2}\sin^{2}\beta$.
For our purpose, we take the forms of $\textbf{Y}_{\nu}$ and $\textbf{M}_R$ in the SSM to be the same forms given in Eq.~(\ref{meff1}).

\section{RG improved effective neutrino mass matrix }

Considering the RG effects in $Y_{\nu}$ and $M_{R}$, we obtain a RG
improved effective neutrino mass matrix via the seesaw
reconstruction at the decoupling scales of the heavy neutrino
singlets. The effective mass matrix at low energies can be found by
running $m_{\rm eff}(Q)$ from the decoupling scales to the
electroweak scale $m_{Z}$. The procedures of the RG running and the
useful formulae are given in Appendix. However, in the case of the
normal hierarchical neutrino mass spectrum of light neutrinos, the
RG running effects from the decoupling scales of heavy Majorana
neutrinos to the electroweak scale  are turned out to be negligibly
small. Ignoring the RG running effects from seesaw scale to the
electroweak scale in Eqs.~(\ref{seesaw1},\ref{seesaw11}), we can obtain the RG improved effective neutrino mass matrix at
low energy  presented as follows,
 \begin{eqnarray}
  m_{\rm eff}\simeq m_{0}\left(\begin{array}{ccc}
  m_{ee}  &  m_{e\mu}  & m^{\ast}_{e\mu}(1+\tilde{a}) \\
  m_{e\mu} &  m_{\mu\mu} & m_{\mu\tau}(1+\tilde{a})  \\
  m^{\ast}_{e\mu}(1+\tilde{a})  &  m_{\mu\tau}(1+\tilde{a})  & m^{\ast}_{\mu\mu}(1+2\tilde{a}) \\
 \end{array}
 \right)+{\cal{O}}(y^{4}_{\tau}) ,
 \label{cmatrix}
 \end{eqnarray}
where the parameter $\tilde{a}$ corresponding to the RG correction
is given as
 \begin{eqnarray}
  \tilde{a} =\left\{
   \begin{array}{ll}
     -\frac{3}{2}y^{2}_{\tau}\cdot t,~~~ & \hbox{\text{for SM},} \\
     y^{2}_{\tau}\cdot t,~~~ & \hbox{\text{for SSM}.}
   \end{array}\right.
 \end{eqnarray}
We recast Eq.~(\ref{cmatrix}) with  the transformation
 $\nu_{\mu}\rightarrow e^{-i\frac{\Psi}{2}}\nu_{\mu}$ and $\nu_{\tau}\rightarrow e^{i\frac{\Psi}{2}}\nu_{\tau}$,
 as \footnote{This effective mass matrix is similar to that of Ref.~\cite{Ahns1}.}
  \begin{eqnarray}
  m_{\rm eff}\simeq m_{0}\left(\begin{array}{ccc}
  m_{ee}  &  \delta_{e\mu}e^{i\Phi}  & \delta_{e\mu}e^{-i\Phi}(1+\tilde{a}) \\
  \delta_{e\mu}e^{i\Phi} &  \delta_{\mu\mu} & m_{\mu\tau}(1+\tilde{a})  \\
  \delta_{e\mu}e^{-i\Phi}(1+\tilde{a})  &  m_{\mu\tau}(1+\tilde{a})  & \delta_{\mu\mu}(1+2\tilde{a}) \\
 \end{array}
 \right)+{\cal{O}}(y^{4}_{\tau}) ,
 \label{cmatrix11}
 \end{eqnarray}
 where $\Phi\equiv\Omega-\frac{\Psi}{2}$ and
  \begin{eqnarray}
  \delta_{e\mu} &=& \sqrt{\lambda^{2}\omega^{2}+\kappa^{2}\chi^{2}+2\lambda\chi\kappa\omega\cos\Delta\phi_{12}},~~~~~~
  \cos\Omega=\frac{\lambda\omega\cos\phi_{1}+\kappa\chi\cos\phi_{2}}{\sqrt{\lambda^{2}\omega^{2}+\kappa^{2}\chi^{2}
  +2\lambda\chi\kappa\omega\cos\Delta\phi_{12}}},\nonumber\\
  \delta_{\mu\mu} &\simeq& \sqrt{\kappa^{4}+\omega^{4}+2\kappa^{2}\omega^{2}\cos2\Delta\phi_{12}}+{\cal{O}}(\frac{1}{\eta}),~~
  \cos\Psi\simeq\frac{\omega^{2}\cos2\phi_{1}+\kappa^{2}\cos2\phi_{2}}{\sqrt{\kappa^{4}+\omega^{4}+2\kappa^{2}\omega^{2}\cos2\Delta\phi_{12}}}
  +{\cal{O}}(\frac{1}{\eta})\nonumber.
 \end{eqnarray}
 It is quite interesting to notice that  the size of $\tilde{a}$ is very small, which breaks the $\mu-\tau$ reflection symmetry
 slightly in the mass matrix $m_{\rm eff}$, however, this tiny breaking effects can play a crucial role in linking between low energy neutrino data
 and leptogenesis, as will be shown later.
This neutrino mass matrix is diagonalized by the PMNS mixing matrix
$U_{\rm PMNS}$, $U^{T}_{\rm PMNS}m_{\rm eff}U_{\rm PMNS}={\rm
Diag}[m_1,m_2,m_3]$, where $m_{i}~ (i=1,2,3)$ indicates the mass
eigenvalues of light Majorana neutrinos.
As a result of the RG effects, CP violation phase $\delta_{\rm CP}$ is shifted by a tiny amount, $\Delta_{\delta}$:
 \begin{eqnarray}
  \delta_{\rm CP}-\frac{\pi}{2}\equiv\Delta_{\delta}\simeq-\tilde{a}\frac{\kappa^{2}+\omega^{2}}{\chi^{2}+\lambda^{2}}\frac{\cos\Phi}{\sin\Phi}~.
  \label{deltaCP1}
 \end{eqnarray}
Here if the exact $\mu-\tau$ reflection symmetry is recovered, then  $\tilde{a}\rightarrow 0$ and $\Delta_{\delta}$ goes to zero.
Atmospheric
mixing angle $\theta_{23}$ is also deviated from $\pi/4$ due to the
RG corrections, and the shift is approximately written by
 \begin{eqnarray}
  \theta_{23}-\frac{\pi}{4}\simeq-2\tilde{a}\frac{\kappa^{2}+\omega^{2}}{\lambda^{2}+\chi^{2}}\simeq2\Delta_{\delta}\tan\Phi~.
  \label{atm1}
 \end{eqnarray}

The magnitude of the unknown angle $\theta_{13}$, which will
complete our knowledge of neutrino mixing, can be written as
 \begin{eqnarray}
  \tan 2\theta_{13}\simeq\frac{\sqrt{2}(\kappa\chi+\lambda\omega)(\chi^{2}+\lambda^{2})}{2(\kappa^2+\omega^2)^2}\sin\Phi.
 \label{theta13}
 \end{eqnarray}
 Note here that $\sin\Phi\simeq\sqrt{\frac{2(\lambda^2+\chi^2)}{\kappa^2+\omega^2}}\delta^{I}_{\mu\tau}$ can be directly linked with leptogenesis through $\delta^{I}_{\mu\tau}$ defined in Eq.~(\ref{leptoComp}).
 Solar neutrino mixing is also presented by
  \begin{eqnarray}
  \tan2\theta_{12}\simeq\frac{\sqrt{2}(\kappa\chi+\omega\lambda)}{2(\kappa^2+\omega^2)}\cos\Phi.
  \label{sol1}
 \end{eqnarray}
 The ratio between $\tan 2\theta_{13}$ and $\tan2\theta_{12}$ can be approximated by the exact $\mu-\tau$ reflection parameter as
  \begin{eqnarray}
  \frac{\tan 2\theta_{13}}{\tan2\theta_{12}}\simeq\frac{\chi^2+\lambda^2}{(\kappa^2+\omega^2)}\tan\Phi,
 \end{eqnarray}
 which is independent of the RG correction.

{}From Eqs.~(\ref{deltaCP1}-\ref{sol1}),
the Jarlskog invariant $J_{\rm CP}$  can be approximated for $\theta_{13}\ll 1$ by
  \begin{eqnarray}
  J_{\rm CP} \simeq \frac{\sin 2\theta_{13}}{8}\sin2\theta_{12}\simeq\frac{\delta^{I}_{\mu\tau}(\lambda^2+\chi^2)}{16(\kappa^2+\omega^2)^4}\{\lambda\chi(\kappa^2-\omega^2)+\kappa\omega(\lambda^2-\chi^2)\cos\Delta\phi_{12}\},
  \label{CP1}
 \end{eqnarray}
 indicating that $J_{\rm CP}$ mainly depends on $\theta_{12}$ and $\theta_{13}$, which is in turn proportional to
 \begin{eqnarray}
  \sin2\Phi\simeq\frac{\delta^{I}_{\mu\tau}}{\delta^{2}_{e\mu}m_{\mu\tau}}\{\lambda\chi(\kappa^2-\omega^2)+\kappa\omega(\lambda^2-\chi^2)\cos\Delta\phi_{12}\}
 \end{eqnarray}
  defined in Eq.~(\ref{cmatrix11}).

\section{Radiatively Induced resonant leptogenesis}

\subsection{In case of the SM}

Let us consider the CP asymmetry generated by the decays of the
heavy Majorana neutrinos $N_{i}$ (i=1, 2).
 In a basis where the right-handed Majorana neutrino mass
matrix is real and diagonal, the CP asymmetry generated through the
interference between tree and one-loop diagrams for the decay of the
heavy Majorana neutrino $N_{i}$ is given, for each lepton flavor
$\alpha~(=e,\mu,\tau)$, by \cite{lepto2,Flavor}
 \begin{eqnarray}\nonumber
  \varepsilon^{\alpha}_{i} &=& \frac{\Gamma(N_{i}\rightarrow l_{\alpha}\varphi)
  -\Gamma(N_{i}\rightarrow \overline{l}_{\alpha}\varphi^{\dag})}{\sum_{\alpha}[\Gamma(N_{i}\rightarrow l_{\alpha}\varphi)
  +\Gamma(N_{i}\rightarrow\overline{l}_{\alpha}\varphi^{\dag})]}\\
  &=& \frac{1}{8\pi(Y_{\nu}Y^{\dag}_{\nu})_{ii}}\sum_{j\neq i}{\rm
  Im}\Big\{(Y_{\nu}Y^{\dag}_{\nu})_{ij}(Y_{\nu})_{i\alpha}(Y_{\nu})^{\ast}_{j\alpha}\Big\}g\Big(\frac{M^{2}_{j}}{M^{2}_{i}}\Big),
 \label{cpasym1}
 \end{eqnarray}
where the function $g(x)$ is given by
 \begin{eqnarray}
  g(x)&=& \sqrt{x}\Big[\frac{1}{1-x}+1-(1+x){\rm ln}\frac{1+x}{x}\Big]~.
  \label{decayfunction}
  \end{eqnarray}
Here $i$ denotes a generation index
and $\Gamma(N_i \rightarrow \cdot \cdot \cdot)$ is the decay width of the $i$th-generation
right-handed neutrino.
 Note here that the flavor effects generated due to the term $ {\rm
  Im}\Big\{(Y_{\nu}Y^{\dag}_{\nu})_{ij}(Y_{\nu})_{i\alpha}(Y_{\nu})^{\ast}_{j\alpha}\Big\}$ disappears in the summation of $\varepsilon_i^{\alpha}$
  for all flavors.
In order for $\varepsilon^{\alpha}_{i}$ to be non-vanishing , not only breaking of the degeneracy of right-handed neutrinos
 but also  non-vanishing ${\rm
Im}[(Y_{\nu}Y^{\dag}_{\nu})_{ik}]$ and/or ${\rm
Re}[(Y_{\nu}Y^{\dag}_{\nu})_{ik}]$ are required at a seesaw scale $M$.

In our scenario, we take the mass hierarchy $M_3\gg M_1 \simeq M_2$,
so that the lepton asymmetry required for a successful leptogenesis is
generated from the decays of both $N_1$ and $N_2$. When two lighter
heavy Majorana neutrinos are nearly degenerate, the dominant
contributions to $\varepsilon^{\alpha}_{i(=1,2)}$ are arisen from
self-energy diagrams and can be written by
 \cite{lepto2}
  \begin{eqnarray}
   \varepsilon_{i}^\alpha \simeq
   \frac{{\rm Im}[\widetilde{H}_{ij}(\widetilde{Y}_\nu)_{i\alpha}(\widetilde{Y}_\nu)^\ast_{j\alpha}]}{16\pi (\widetilde{Y}_{\nu} \widetilde{Y}^{\dag}_{\nu})_{ii}\delta^{ij}_{N}}\Big (1+\frac{\Gamma_{j}^2}{4M^{2}_{Ri} \delta^{ij2}_{N}} \Big)^{-1}~,~~~~(i\neq j)
   \label{FlaLepto}
  \end{eqnarray}
where $\widetilde{H}_{ij}, \widetilde{Y}_\nu$ and $\delta^{ij}_{N}$ are defined
in Eqs.~(\ref{deltaM},\ref{seesaw1},\ref{seesaw2},\ref{ImReH'}) of Appendix, and the term containing  decay width is negligibly  small.
We notice
from Eq.~(\ref{FlaLepto}) that $\varepsilon^{\alpha}_{i}$ is resonantly enhanced when $\Gamma_{j}\simeq(M^{2}_{Ri}-M^{2}_{Rj})/M_{Ri}$.
Here, the RG evolution of the parameter $\delta^{ij}_{N}$ reflecting the mass splitting of the degenerate heavy Majorana neutrinos is governed by Eq.~(\ref{degeneracy1}). In the limit $\delta^{ij}_{N}\ll1$, the leading-log approximation for
$\delta^{jk}_{N}$ can be easily found to be
 \begin{eqnarray}
   \delta^{ij}_{N}\simeq2[\widetilde{H}_{ii}-\widetilde{H}_{jj}]\cdot t~,
   \label{deltaN}
 \end{eqnarray}
 where the energy scale parameter $t$ is defined by Eq.~(\ref{scalet}).
 By using Eqs.~(\ref{FlaLepto},\ref{deltaM}), $\varepsilon_{i}^{\alpha}$ can be expressed as
 \begin{eqnarray}
   \varepsilon_{1(2)}^e
   &\simeq& \pm \frac{\delta^{I}_{\mu\tau}\epsilon}{16\pi\cdot h_{1(2)}}
   \Big\{\frac{\delta_{ee}}{\delta^{R}_{\mu\tau}}-\epsilon\frac{\delta'_{ee}}{\delta^{R}_{\mu\tau}}\cdot t\Big\}+{\cal{O}}(t^{2})~,\nonumber\\
   \varepsilon_{1(2)}^{\mu}
   &\simeq& \pm \frac{\delta^{I}_{\mu\tau}\epsilon}{32\pi\cdot h_{1(2)}}
   \Big\{1-2\epsilon\frac{\delta'_{\mu\tau}}{\delta^{R}_{\mu\tau}}\cdot t\Big\}+{\cal{O}}(t^{2})~,\nonumber\\
   \varepsilon_{1(2)}^{\tau}
   &\simeq& \pm \frac{3\delta^{I}_{\mu\tau}\epsilon}{32\pi\cdot h_{1(2)}}
   \Big\{1-\big(\frac{2}{3}\epsilon\frac{\delta'_{\mu\tau}}{\delta^{R}_{\mu\tau}}+3y^{2}_{\tau}\big)\cdot t\Big\}+{\cal{O}}(t^{2}),
  \label{FlaLepto1}
 \end{eqnarray}
where
 \begin{eqnarray}
  \delta_{ee} &\equiv& \frac{\chi^{2}-\lambda^{2}}{2}\sin2\alpha+\lambda\chi\cos2\alpha~,~~~~~\delta'_{ee} \equiv (\chi^{2}-\lambda^{2})\cos2\alpha-2\lambda\chi\sin2\alpha~,\nonumber\\
  \delta^{R}_{\mu\tau} &\equiv& \frac{\kappa^{2}-\omega^{2}}{2} \sin2\alpha+
  \kappa\omega\cos2\alpha\cos\Delta\phi_{12}~,\nonumber\\
  \delta^{I}_{\mu\tau} &\equiv& \kappa\omega\sin\Delta\phi_{12}~,~~~~~~~~~~~~
  \delta'_{\mu\tau} \equiv (\kappa^{2}-\omega^{2})\cos2\alpha-2\kappa\omega\sin2\alpha\cos\Delta\phi_{12}~,
  \label{leptoComp}
 \end{eqnarray}
the parameter $\epsilon$ presenting RG corrections is defined in (\ref{RGparameter}) and the two parameters $h_{1(2)}$ are defined as
 \begin{eqnarray}
  h_{1} &=& \widetilde{H}_{11}/b^{2}_{3}\simeq
  (2\kappa^{2}+\chi^{2})\sin^{2}\alpha+(\lambda^{2}+2\omega^{2})\cos^{2}\alpha+\{\lambda\chi+2\kappa\omega\cos\Delta\phi_{12}\}\sin2\alpha~,\nonumber\\
  h_{2} &=& \widetilde{H}_{22}/b^{2}_{3}\simeq
  (2\kappa^{2}+\chi^{2})\cos^{2}\alpha+(\lambda^{2}+2\omega^{2})\sin^{2}\alpha-\{\lambda\chi+2\kappa\omega\cos\Delta\phi_{12}\}\sin2\alpha~.
 \label{h}
 \end{eqnarray}

Note here that $\delta_{ee}/\delta^{R}_{\mu\tau}=-2$, which is justified from Eq.~(\ref{tan2alpha}),
and the sign of plus and minus in Eq.~(\ref{FlaLepto1}) correspond to the case of the decay of $N_{1}$ and $N_{2}$, respectively.
It is worthwhile to notice that since the angle $\alpha$ given by Eq.~(\ref{tan2alpha}) is limited by $-45^0<\alpha<45^0$,
$h_{2}$ (order of 10) is always greater than $h_{1}$ (order of 1), as will be shown later. This implies that $\varepsilon^{\alpha}_{1}$ is
dominant over $\varepsilon^{\alpha}_{2}$ because of $h_{2}\gg
h_{1}$. Since the total CP asymmetries $\varepsilon_{i}=\sum_{\alpha}\varepsilon^{\alpha}_{i}$ are obtained
by summing $\varepsilon_{i}^{\alpha}$ over the lepton flavors $\alpha$,
with the help of Eqs.~(\ref{leptoComp},\ref{tan2alpha},\ref{deltaM},\ref{RGparameter}), we can express the
lepton asymmetries as follows \footnote{Eq.~(\ref{convLepto2}) can also be obtained directly by using Eq.~(\ref{ImReH'}).}
 \begin{eqnarray}
  \varepsilon_{1(2)}
  &\simeq&  \mp \frac{3\delta^{I}_{\mu\tau}\epsilon}{16\pi h_{1(2)}}y_\tau^2\cdot t~.
 \label{convLepto2}
 \end{eqnarray}
 where the sign of minus and plus correspond to the case of the decay of $N_{1}$ and $N_{2}$, respectively.
 As indicated in Eq.~(\ref{FlaLepto}), the CP asymmetries $\varepsilon^{\alpha}_{i}$ weakly depend on the heavy Majorana neutrino scale $M$.
{}From Eqs.~(\ref{FlaLepto1},\ref{convLepto2}), we see that $\varepsilon_{i}\propto y^{4}_{\tau} t$ and
 $\varepsilon^{\alpha}_{i}\propto y^{2}_{\tau}$. Therefore, the CP asymmetry $\varepsilon^{e}_{1}$ gets enhanced by
 $\varepsilon^{e}_{1}/\varepsilon_{1}\sim 2/(3y^{2}_{\tau}t)$ due to flavor effects \cite{Ahn:2007mj}.

Below temperature $T\sim M_{i}\lesssim10^{9}$ GeV, it is known that
muon and tau charged lepton Yukawa interactions are much faster than
the Hubble expansion parameter rendering the $\mu$ and $\tau$ Yukawa
couplings in equilibrium.
Then, the processes which wash out lepton number are flavor dependent and thus
the lepton asymmetries for each flavor should be treated separately with different wash-out factors.
Once the initial values of $\varepsilon^{\alpha}_{i}$ are fixed, the final result of $\eta_{B}$ or $Y_{B}$ can be obtained by solving a set of
flavor-dependent Boltzmann equations including the decay, inverse decay,
and scattering processes as well as the non-perturbative sphaleron
interaction. In order to estimate the wash-out effects, we introduce
the parameters $K^{\alpha}_{i}$ which are the wash-out factors mainly due to
the inverse decay of the Majorana neutrino $N_{i}$ into the lepton
flavor $\alpha(=e,\mu,\tau)$ \cite{Abada}.
The explicit form of $K^{\alpha}_{i}$ is given by
 \begin{eqnarray}
  K^{\alpha}_{i} =\frac{\Gamma(N_{i}\rightarrow \varphi l_{\alpha})}{H(M_{Ri})}= \frac{\widetilde{m}^{\alpha}_{i}}{m_{\ast}}~,
  ~\text{with}~~ \widetilde{m}^{\alpha}_{i}=(Y^{\dag}_{\nu})_{\alpha
  i}(Y_{\nu})_{i\alpha}\frac{\upsilon^{2}}{M_{Ri}}~,
  \label{K-factor2}
 \end{eqnarray}
 where  $m_{\ast}\simeq10^{-3}$ eV, and $\Gamma(N_{i}\rightarrow \varphi l_{\alpha})$ and $ H(M_{Ri})$ denote the partial decay rate of the process
$N_{i}\rightarrow l_{\alpha}+\varphi^{\dag}$ and the Hubble
parameter at temperature $T\simeq M_{Ri}$, respectively. The decay
rate of $N_{i}$ to the leptons with flavor $l_{\alpha}$ is
parameterized by $\widetilde{m}^{\alpha}_{i}$ and the trace
$\sum_{\alpha}\widetilde{m}^{\alpha}_{i}$ coincides with the
parameter $\widetilde{m}_{i}$.
{}From Eq.~(\ref{K-factor2}) and
Eq.~(\ref{seesaw1}), the wash-out parameters are given as
 \begin{eqnarray}
  K^{e}_{1} &\simeq& \frac{m_{0}}{m_{\ast}}(\chi s_{\alpha}+\lambda c_{\alpha})^{2}~,~~~K^{e}_{2} \simeq \frac{m_{0}}{m_{\ast}}(\chi c_{\alpha}-\lambda s_{\alpha})^{2}~,\nonumber\\
  K^{\mu,\tau}_{1} &\simeq& \frac{m_{0}}{m_{\ast}}(s^{2}_{\alpha}\kappa^{2}+c^{2}_{\alpha}\omega^{2}
  +2\kappa\omega c_{\alpha}s_{\alpha}\cos\Delta\phi_{12})~,\nonumber\\
  K^{\mu,\tau}_{2} &\simeq& \frac{m_{0}}{m_{\ast}}(c^{2}_{\alpha}\kappa^{2}+s^{2}_{\alpha}\omega^{2}
  -2\kappa\omega c_{\alpha}s_{\alpha}\cos\Delta\phi_{12})~.
  \label{K-eM}
  \end{eqnarray}
The final baryon asymmetry $Y_B$ is then
given by \cite{Abada}
 \begin{eqnarray}
  Y_{\emph{B}}&\simeq&
  \frac{12}{37}\sum_{N_{i}}\Big[Y^{e}_{i}\Big(\varepsilon^{e}_{i},\frac{151}{179}\widetilde{m}^{e}_{i}\Big)
  +Y^{\mu}_{i}\Big(\varepsilon^{\mu}_{i},\frac{344}{537}\widetilde{m}^{\mu}_{i}\Big)
  +Y^{\tau}_{i}\Big(\varepsilon^{\tau}_{i},\frac{344}{537}\widetilde{m}^{\tau}_{i}\Big)\Big]~.
 \label{baryon3}
 \end{eqnarray}

Notice that each CP asymmetry for a single flavor given in Eq.~(\ref{baryon3}) is weighted differently by the corresponding
wash-out parameter given by Eq.~(\ref{K-factor2}), and appears with different weight in the final formula for the baryon
asymmetry~\cite{Abada}. {}From our numerical estimate, we found that the wash-out factors $K_{i}^{\alpha}$ except for $K^{e}_{2}$ are much
greater than one indicating that corresponding lepton flavor asymmetries are strongly washed out,
which is undesirable for successful leptogenesis.
Since $K_2^{e}$ can be much less than one and the lepton asymmetries given in Eq.~(\ref{FlaLepto1})
are the same order of magnitude, the lepton asymmetry for electron flavor in the weak wash-out regime
leads to the dominant contribution to the lepton asymmetry, which can be
enough to give rise to successful baryogenesis. Thus, $Y_B$ is
approximately given by
\begin{eqnarray}
  Y_{\emph{B}} &\simeq&
  \frac{12}{37}Y^{e}_{2}\Big(\varepsilon^{e}_{2},\frac{151}{179}\widetilde{m}^{e}_{2}\Big)~,
 \end{eqnarray}
and the magnitudes of $K_{2} (=K^e_2+K^{\mu}_2+K^{\tau}_2)$ and
$K^{e}_{2}$ are approximately
 \begin{eqnarray}
  K_{2}\sim130-160,~~~~~~K^{e}_{2}\sim10^{-5}-1~.
 \end{eqnarray}
In the weak wash-out regime ($K^{\alpha}_{i}<1$), the lepton
asymmetry $Y_2^e$ generated through the decay of $N_{2}$ is given
by~\cite{Abada}
  \begin{eqnarray}
  Y^{e}_{2} &\simeq&
  1.5\frac{\varepsilon^{e}_{2}}{g_{\ast}}\Big(\frac{\widetilde{m}_{2}}{3.3\times10^{-3} eV}\Big)
  \Big(\frac{\widetilde{m}^{e}_{2}}{3.3\times10^{-3} eV}\Big).
  \label{weak}
 \end{eqnarray}
Therefore, the resulting baryon-to-photon ratio $\eta^{\rm f}_{B}$
can be simply given as
 \begin{eqnarray}
  \eta^{\rm f}_{B}\simeq -10^{-3}\cdot\varepsilon^{e}_{2}K^{e}_{2}K_{2}~.
  \label{baryonasym}
 \end{eqnarray}
As will be shown later, the wash-out factor $K^{e}_2$ is proportional to the low energy mixing  angle $\theta_{13}$ and there is a
connection between $\eta^{\rm f}_{B}$ and low energy neutrino observables such as $\theta_{13}$ and $J_{\rm CP}$.

On the other hand, the ratio of $\eta_B$ without lepton flavor
effects compared to $\eta^{\rm f}_B$ with lepton flavor effects
\cite{Ahn:2007mj} is presented by
  \begin{eqnarray}
   \frac{\eta_{B}}{\eta^{\rm
   f}_{B}}\sim\frac{\varepsilon_{2}}{\varepsilon^{e}_{2}}\frac{1}{K^{e}_{2}(K_{2})^2}\approx\frac{3y^{2}_{\tau}}{32\pi^{2}}{\rm
   ln}\big(\frac{M}{M_{\rm GUT}}\big)\frac{1}{K^{e}_{2}(K_{2})^2}~.
  \label{ratioBaryon}
  \end{eqnarray}
 Thus, without taking lepton flavor effects into account,
 the prediction of $\eta_B$ is suppressed by $\sim10^{-8}/K^{e}_{2}$, {\it i.e.,} $4\sim 8$ orders of magnitude
 compared with $\eta^{\rm f}_B$, which is too small to give  a successful leptogenesis.

\subsection{In case of the SSM}

Using the same way as in the case of the SM,  we obtain from
Eqs. (\ref{cpasym1},\ref{deltaN}) the lepton asymmetries for each lepton flavor
 expressed in terms of low energy parameters as
\begin{eqnarray}
   \varepsilon_{1(2)}^{e}
   &\simeq& \mp \frac{\delta^{I}_{\mu\tau}\epsilon}{48\pi h_{1(2)}}\Big\{\frac{\delta_{ee}}{\delta^{R}_{\mu\tau}}+\frac{2\epsilon}{3}\frac{\delta'_{ee}}{\delta^{R}_{\mu\tau}}\cdot t\Big\}+{\cal{O}}(t^{2})~,\nonumber\\
   \varepsilon_{1(2)}^{\mu}
   &\simeq& \mp \frac{\delta^{I}_{\mu\tau}\epsilon}{96\pi h_{1(2)}}\Big\{1+\frac{4\epsilon}{3}\frac{\delta'_{\mu\tau}}{\delta^{R}_{\mu\tau}}\cdot t\Big\}+{\cal{O}}(t^{2})~,\nonumber\\
   \varepsilon_{1(2)}^{\tau}
   &\simeq& \mp \frac{\delta^{I}_{\mu\tau}\epsilon}{32\pi h_{1(2)}}\Big\{1+2\big(y^{2}_{\tau}+\frac{2\epsilon}{9}\frac{\delta'_{\mu\tau}}{\delta^{R}_{\mu\tau}}\big)\cdot t\Big\}+{\cal{O}}(t^{2})~,
 \label{FlaLepto1M}
 \end{eqnarray}
where the sign of minus and plus correspond to the case of the decay of $N_{1}$ and $N_{2}$, respectively.
Here, we notice that the lepton asymmetry $\varepsilon^\alpha_{1,2}$ is proportional to the parameter $\epsilon$,
and at the same time can be enhanced by $\tan^{2}\beta$, because $\epsilon\propto y^{2}_{\tau}=y^{2}_{\tau\rm SM}\tan^2\beta$
as shown in Eq.~(\ref{RGparameter}). Here, we also note that in the SSM  the loop function $g(x)$ in Eq.~(\ref{cpasym1})
and the mass splitting
$\delta^{ij}_{N}$  are give by
 \begin{eqnarray}
  g(x) = \sqrt{x}\Big[\frac{2}{1-x}-{\rm ln}\frac{1+x}{x}\Big]~,~~~~~~~~~~\delta^{ij}_{N}\simeq4[\widetilde{H}_{ii}-\widetilde{H}_{jj}]\cdot t~,
  \label{decayfunctionM}
  \end{eqnarray}
which are different from those in the SM. Summing over the lepton flavors $\alpha$ in Eq.~(\ref{cpasym1}),  the total lepton
asymmetries are approximately given by
 \begin{eqnarray}
  \varepsilon_{1(2)}
  &\simeq& \mp \frac{\delta^{I}_{\mu\tau}\epsilon}{24\pi h_{1(2)}}y_\tau^2\cdot t~.
 \label{convLeptoM2}
 \end{eqnarray}

It is worthwhile to notice that the radiatively induced lepton
asymmetries $\varepsilon_{1(2)}$ are  proportional to
$y^{4}_{\tau}=y^{4}_{\tau\rm SM}(1+\tan^{2}\beta)^{2}$, and thus for
large $\tan\beta$ it can be highly enhanced and proportional to
$\tan^{4}\beta$. Furthermore, it has an explicit dependence of the
evolution scale $t$. These two points are already mentioned in Ref.
\cite{Ahn:2006rn}.
Taking lepton flavor effects into account, in this case the final baryon asymmetry is given  as \cite{Abada}
 \begin{eqnarray}
  Y_{\emph{B}} \simeq
  \frac{10}{31}\sum_{N_{i}}\Big[Y^{e}_{i}\Big(\varepsilon^{e}_{i},\frac{93}{110}\widetilde{m}^{e}_{i}\Big)
  +Y^{\mu}_{i}\Big(\varepsilon^{\mu}_{i},\frac{19}{30}\widetilde{m}^{\mu}_{i}\Big)
  +Y^{\tau}_{i}\Big(\varepsilon^{\tau}_{i},\frac{19}{30}\widetilde{m}^{\tau}_{i}\Big)\Big]~.
 \label{baryon3M}
 \end{eqnarray}
Similar to the SM, below temperatures $T\sim M_{i}\lesssim(1+\tan^{2}\beta)10^{9}$, muon and
tau charged lepton Yukawa interactions are much faster than the
Hubble expansion parameter rendering the $\mu$ and $\tau$ Yukawa
couplings in equilibrium.
To obtain the final baryon asymmetry survived, we should consider the wash-out factors.
Similar to the SM, the magnitudes of $K_i^{\alpha}$ except for
$K_2^e$ are greater than one in our scenario. So, we will consider
two cases depending on the magnitude of $K_2^e$ :
(i)$K^{e}_{2}\lesssim1$ and
$K^{\mu,\tau}_{2}>1~\text{and}~K^{\alpha}_{1}>1$,
(ii) $K^{\alpha}_{1(2)}>1$.

\subsubsection{\bf In case of $K^{e}_{2}\lesssim1$ and $K^{\mu,\tau}_{2}>1~\text{and}~K^{\alpha}_{1}>1~(\alpha=e,\mu,\tau)$}

This case implies that the muon and tau lepton asymmetries are
strongly washed out, whereas the electron asymmetry generated
through the decay of $N_2$ is weakly washed out. The form of $Y_2^e$
in the weak wash-out regime in the SSM is the same as that in the SM
given by Eq.~(\ref{weak}). In this case, similar to the SM, the
contribution of $Y_2^e$ to $Y_B$ is dominant over the others. Thus,
$Y_B$ is approximately give by
 \begin{eqnarray}
  Y_{\emph{B}} \simeq
  \frac{10}{31}Y^{e}_{2}\Big(\varepsilon^{e}_{2},\frac{93}{110}\widetilde{m}^{e}_{2}\Big)~,
 \end{eqnarray}
and the magnitudes of $K_{2}$ and $K^{e}_{2}$ are approximately
 \begin{eqnarray}
  K_{2}\sim80~(\tan\beta=1),~~~\sim40~(\tan\beta=10-60)~,~~~~~~K^{e}_{2}\sim10^{-5}-1~.
 \end{eqnarray}
The resulting baryon-to-photon ratio $\eta^{\rm f}_{B}$ is simply
given as
 \begin{eqnarray}
  \eta^{\rm f}_{B}\simeq -10^{-3}\cdot\varepsilon^{e}_{2}K^{e}_{2}K_{2}~.
  \label{baryonasymSSM}
 \end{eqnarray}
Just like the SM, one can see that there is a connection between
baryon asymmetry and low energy neutrino observables.
The ratio of $\eta_B$ without lepton flavor effects  to $\eta^{\rm f}_B$ with lepton flavor effects are simply given
by \cite{Ahn:2007mj}
  \begin{eqnarray}
   \frac{\eta_{B}}{\eta^{\rm
   f}_{B}}\sim\frac{\varepsilon_{2}}{\varepsilon^{e}_{2}}\frac{1}{K^{e}_{2}(K_{2})^2}\approx\frac{y^{2}_{\tau}}{16\pi^{2}}{\rm
   ln}\big(\frac{M}{M_{\rm GUT}}\big)\frac{1}{K^{e}_{2}(K_{2})^2}~.
  \label{ratioBaryonSSM}
  \end{eqnarray}
{}From this result, we see that without taking lepton flavor effects into account the prediction of $\eta_B$ is suppressed by
$\sim10^{-8}(1+\tan^{2}\beta)/K^{e}_{2}$,  compared with $\eta^{\rm f}_B$ . However, on the contrary to the flavor independent
leptogenesis in the SM, it is possible to get a right amount of baryon asymmetry without flavor effects for the case of large $\tan\beta ~(\sim 60)$.

\subsubsection{\bf In case of $K^{\alpha}_{1,2}>1,~(\alpha=e,\mu,\tau)$}

In the strong wash-out regime
$K^{\alpha}_{1,2}>1~(\alpha=e,\mu,\tau)$, the lepton asymmetry for
the flavor $l_{\alpha}$ generated through the decay of $N_{i}$ is
given by  \cite{Abada}
  \begin{eqnarray}
  Y^{\alpha}_{i} &\simeq&
  0.3\frac{\varepsilon^{\alpha}_{i}}{g_{\ast}}\Big(\frac{0.55\times10^{-3}eV}{\widetilde{m}^{\alpha}_{i}}\Big)^{1.16}.
  \label{strong}
 \end{eqnarray}
In this case, given the initial thermal abundance of $N_{i}$ and the
condition for $K^{\alpha}_{i}$, the baryon asymmetry including
lepton flavor effects is approximately given by \cite{PU2}
  \begin{eqnarray}
   \eta^{\rm f}_{B}\simeq-10^{-2}\sum_{i}\sum_{\alpha}\varepsilon^{\alpha}_{i}\kappa^{\alpha}_{i}~,
   ~~\text{with}~\kappa^{\alpha}_{i}\equiv\frac{K^{\alpha}_{i}}{K_{i}K^{\alpha}}~.
   \label{etaB}
  \end{eqnarray}
Here, we note that each lepton asymmetry $\varepsilon_i^{\alpha}$
are almost the same order of magnitudes, even though
$\varepsilon^{\tau}_{1}$ for large $\tan\beta$ and
$\varepsilon^{e}_{1}$ for small $\tan\beta$ are slightly dominant
within a few factors. According to our numerical estimate, the
typical sizes of the suppression factors $\kappa^{\alpha}_i$ are
 \begin{eqnarray}
  \kappa^{e}_{1}&\sim&10^{-1}>\kappa^{\mu,\tau}_{2}\sim2\times10^{-2}\gtrsim\kappa^{\mu,\tau}_{1}>\kappa^{e}_{2}\sim10^{-2}~~~~~~~\text{for}~\tan\beta\gtrsim3,\nonumber\\
  \kappa^{e}_{1}&\sim&5\times10^{-1}>\kappa^{\mu,\tau}_{2}\sim10^{-2}\gtrsim\kappa^{\mu,\tau}_{1}>\kappa^{e}_{2}\sim4\times10^{-3}~~\text{for}~\tan\beta<3~,
  \label{Kstrong}
 \end{eqnarray}
which indicates that the flavor dependent wash-out factors  equally
contribute to leptogenesis.

Using Eq.~(\ref{Kstrong}), the resulting baryon-to-photon ratio $\eta^{\rm f}_{B}$ can be simply approximated as an order of magnitude
 \begin{eqnarray}
  \eta^{\rm f}_{B}\sim 10^{-11}(1+\tan^{2}\beta)~,
  \label{baryonasymST}
 \end{eqnarray}
which represents that at least $\tan\beta\gtrsim7$ is needed for the flavored leptogenesis to successfully work (see, Fig.~\ref{Fig6} in sec. V).

As a consequence, in the strong wash-out case, the resulting baryon asymmetry is too small to give a successful leptogenesis in the SM,
whereas at least $\tan\beta\gtrsim7$ would be necessary to accommodate required baryon asymmetry in the SSM,
and there exists also a connection between leptogenesis and low energy observable $\theta_{12}$.
Note that the strong wash-out case only allow $\theta_{13}$ to be $\gtrsim6^{\circ}$ which is from the condition $K^{e}_{2}>1$.

\section{Numerical Analysis and Results}

In order to estimate the RG evolutions of neutrino Dirac-Yukawa
matrix and heavy Majorana neutrino masses from the GUT scale to a
seesaw scale, we numerically solve all the relevant RG equations presented
in \cite{Antusch:2005gp}.
 In our numerical calculation of the RG
evolutions, we first fix the values of two masses of heavy Majorana
neutrinos with hierarchy $M_3 \gg M_1=M_2=M$ to be $M_3=10^{12}\;{\rm GeV}$, and
$M_2=5\times10^{6}\;{\rm GeV}$ as inputs\footnote{We note that the
mass $M$ can be as light as $10^3$ GeV in our scenario.}.
Then we solve the RG equations by
varying input values of all the parameter space $\{b_{3},\kappa,\omega,\lambda,\chi,\phi_{1},\phi_{2}\}$ given at the
GUT scale. Then, finally we select the parameter space allowed by
low energy neutrino experimental data. At present, we have five
experimental data, which are taken as low energy constraints in our
numerical analysis, given at $3\sigma$  by \cite{Maltoni:2004ei},
  \begin{eqnarray}
   &&0.26\leq \sin^{2}\theta_{12}\leq0.40~,\quad~~0.34\leq\sin^{2}\theta_{23}\leq0.67~,\quad
   ~~\sin^{2}\theta_{13}\leq0.050,\nonumber\\
   &&2.0\leq\Delta m^{2}_{\rm Atm}[10^{-3} {\rm eV}^{2}]\leq2.8~,\quad\quad
   ~~7.1\leq\Delta m^{2}_{\rm Sol}[10^{-5}{\rm eV}^{2}]\leq8.3,
  \label{exp bound}
  \end{eqnarray}
where $\Delta m^{2}_{\rm Atm}\equiv m^{2}_{3}-m^{2}_{2}$ and $\Delta
m^{2}_{\rm Sol} \equiv m^{2}_{2}-m^{2}_{1}$.
In addition, the current measurement of baryon asymmetry of our universe $\eta_B^{\rm exp}$
we take in our numerical calculation is given by \cite{cmb}
\begin{eqnarray}
\eta_B^{\rm exp}=(6.2 \pm 0.15) \times 10^{-10}. \label{etaBexp}
\end{eqnarray}
Using the
results of the RG evolutions, we estimate baryon asymmetry for the
parameter space constrained from the low energy experimental data.

\subsection{In case of $K^{e}_{2}\lesssim1$ and $K^{\mu,\tau}_{2}>1~\text{and}~K^{\alpha}_{1}>1~(\alpha=e,\mu,\tau)$}


\begin{figure}[t]
\hspace*{-2cm}
\begin{minipage}[t]{6.0cm}
\epsfig{figure=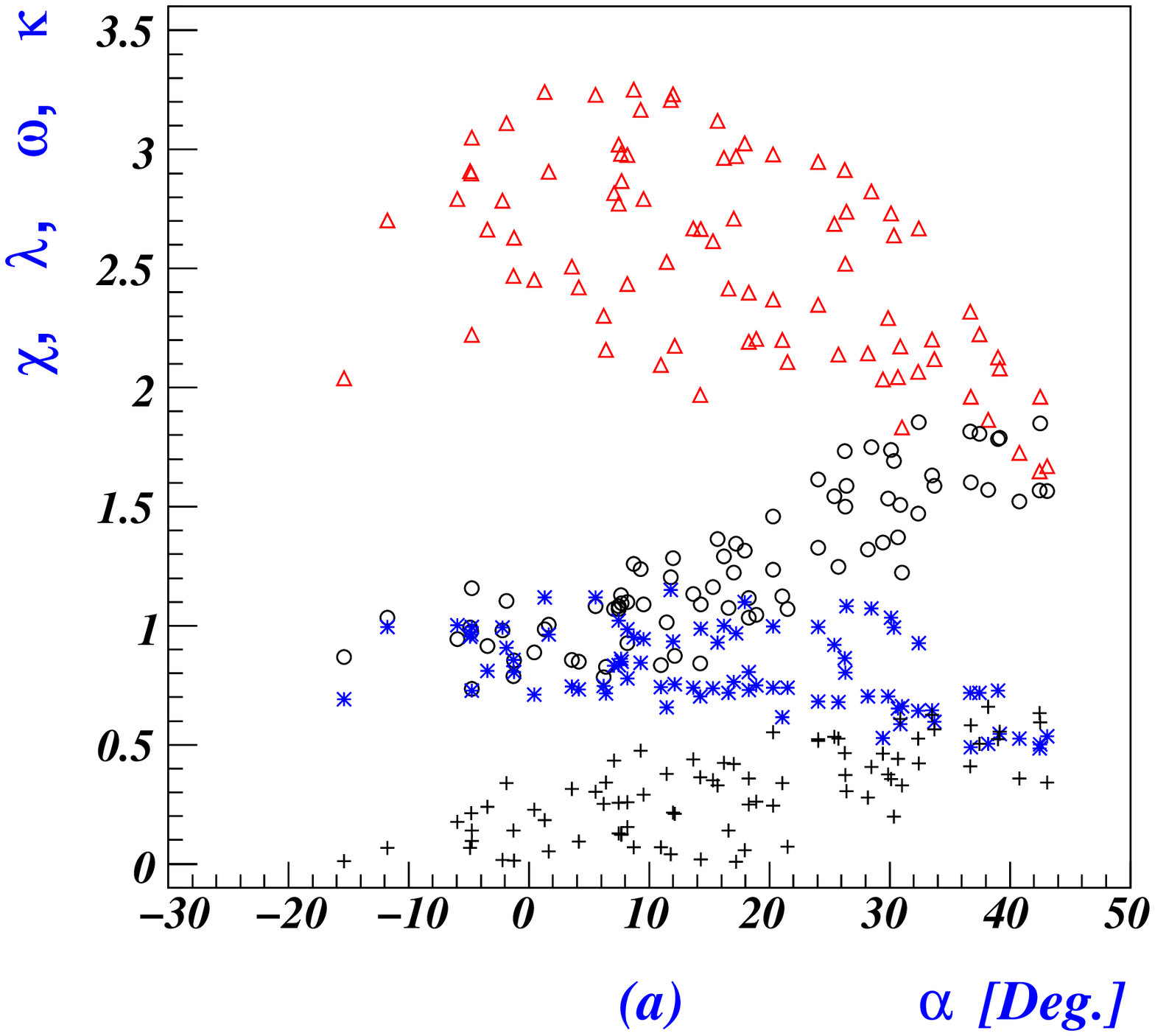,width=6.5cm,angle=0}
\end{minipage}
\hspace*{2.0cm}
\begin{minipage}[t]{6.0cm}
\epsfig{figure=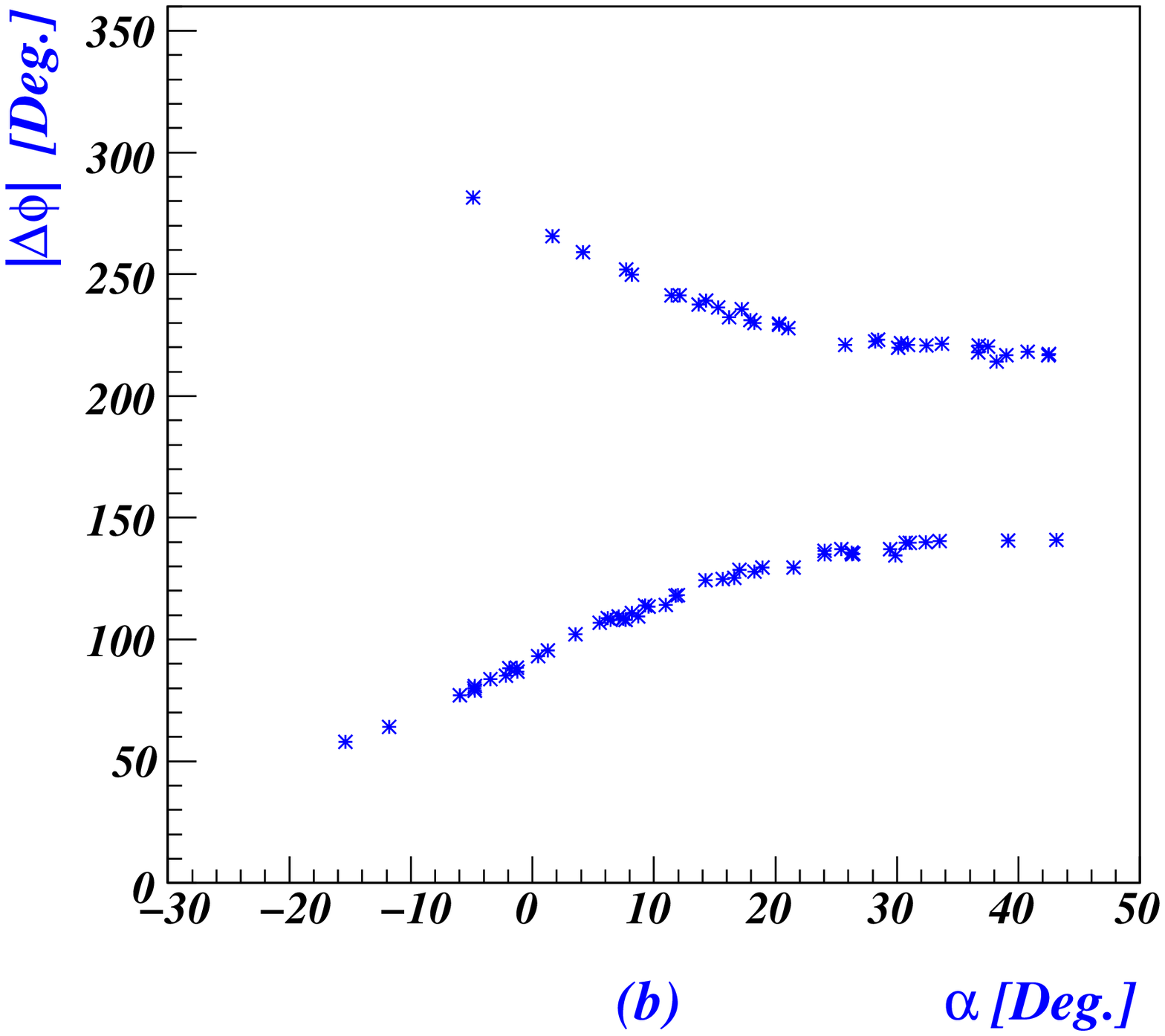,width=6.5cm,angle=0}
\end{minipage}
\vspace*{-1.0cm} \hspace*{-2cm}
\begin{minipage}[t]{6.0cm}
\epsfig{figure=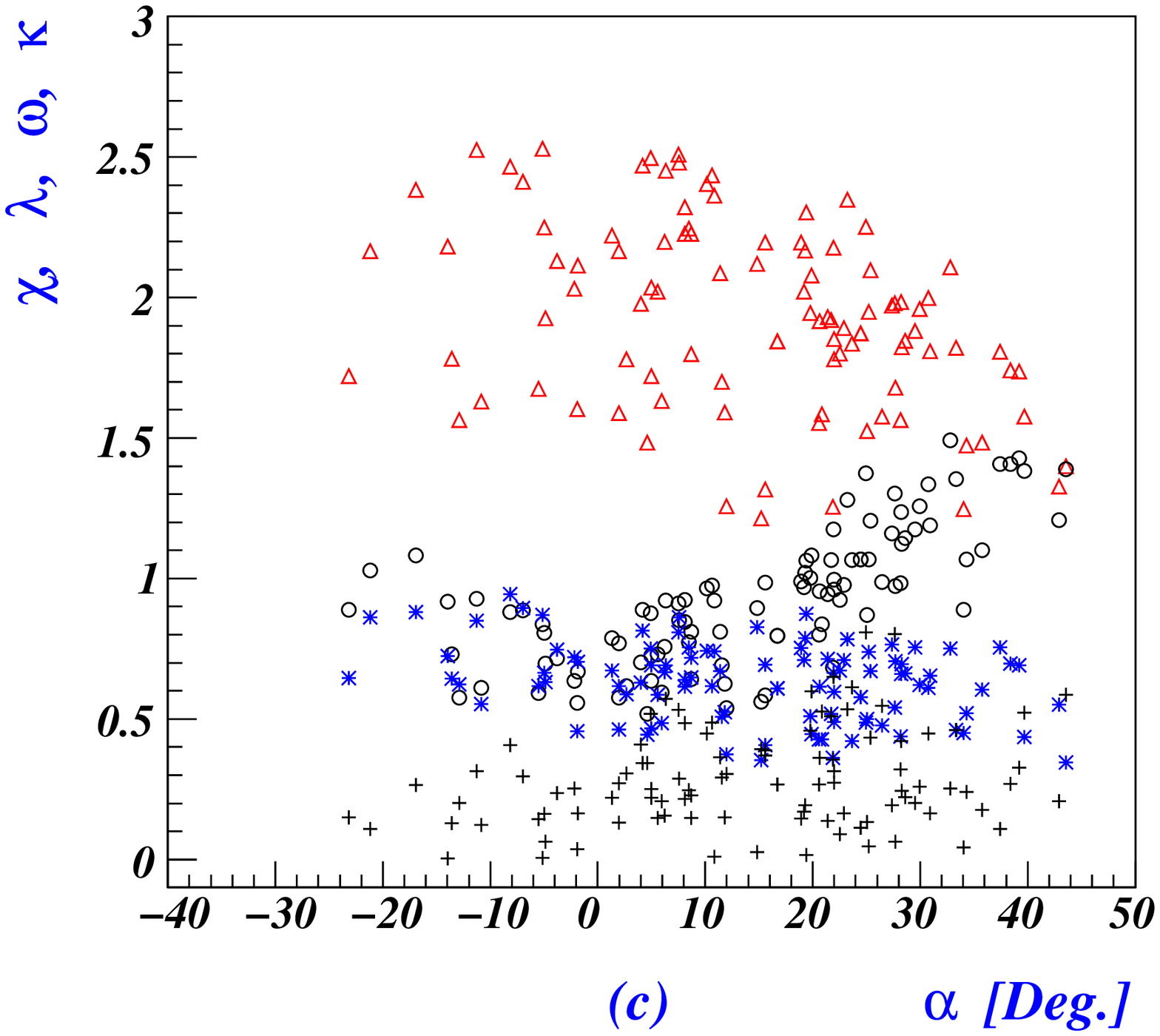,width=6.5cm,angle=0}
\end{minipage}
\hspace*{2.0cm}
\begin{minipage}[t]{6.0cm}
\epsfig{figure=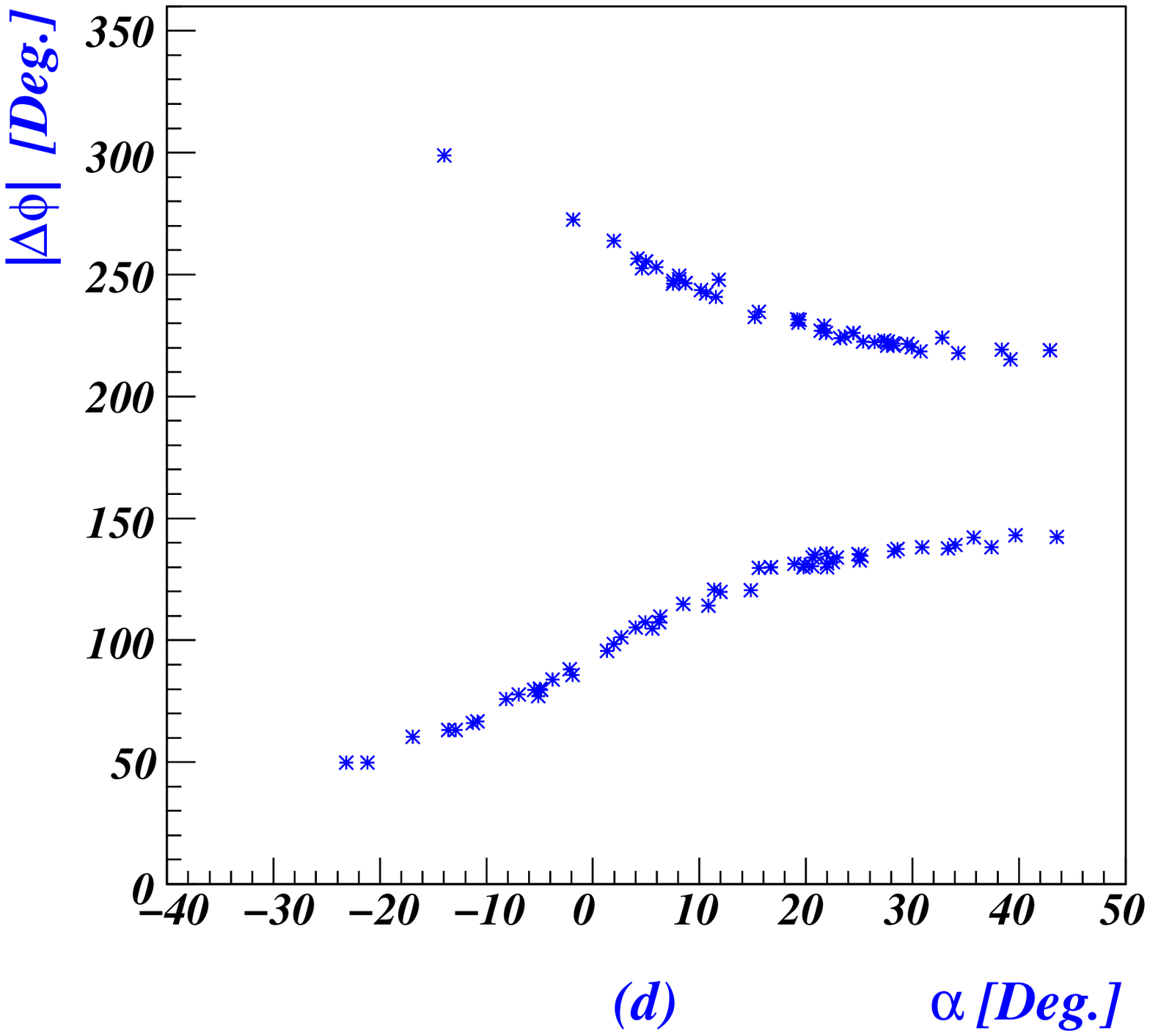,width=6.5cm,angle=0}
\end{minipage}
\vspace*{0.9cm}
\caption{\label{Fig1} Parameter regions allowed by the
experimental constraints at $3\sigma$ C.L. given in Eq.~(\ref{exp bound}) for
$M_3=10^{12}\;{\rm GeV}$, $M_2=5\times10^{6}\;{\rm GeV}$. The
figures in left-hand side exhibit how the parameters $\kappa({\rm triangles})$,
$\omega(\rm circles)$, $\lambda({\rm asters})$ and $\chi({\rm
crosses})$ are correlated with the angle $\alpha$, respectively and
the figures in right-hand side represent how $|\Delta\phi|$ depends on the angle
$\alpha$, where $\Delta\phi=\phi_{2}-\phi_{1}$, among the phases
assigned in the neutrino Dirac-Yukawa matrix given in the form of
Eq.~(\ref{input1}). Upper panels correspond to the case of the SM, whereas
lower panels correspond to the case of the SSM for
$\tan\beta=10$.}
\end{figure}

For this case of wash-out factors, we found that baryogenesis could be successfully implemented in the SM and the SSM through
electron flavor dominant leptogenesis. In addition, we show that the successful leptogenesis could be linked with low energy
observables ($\theta_{13}$ and $J_{\rm CP}$) through the RG parameter $\epsilon$ given in Eq.~(\ref{RGparameter}) as well as $K^{e}_{2}$.

\begin{figure}[t]
\hspace*{-2cm}
\begin{minipage}[t]{6.0cm}
\epsfig{figure=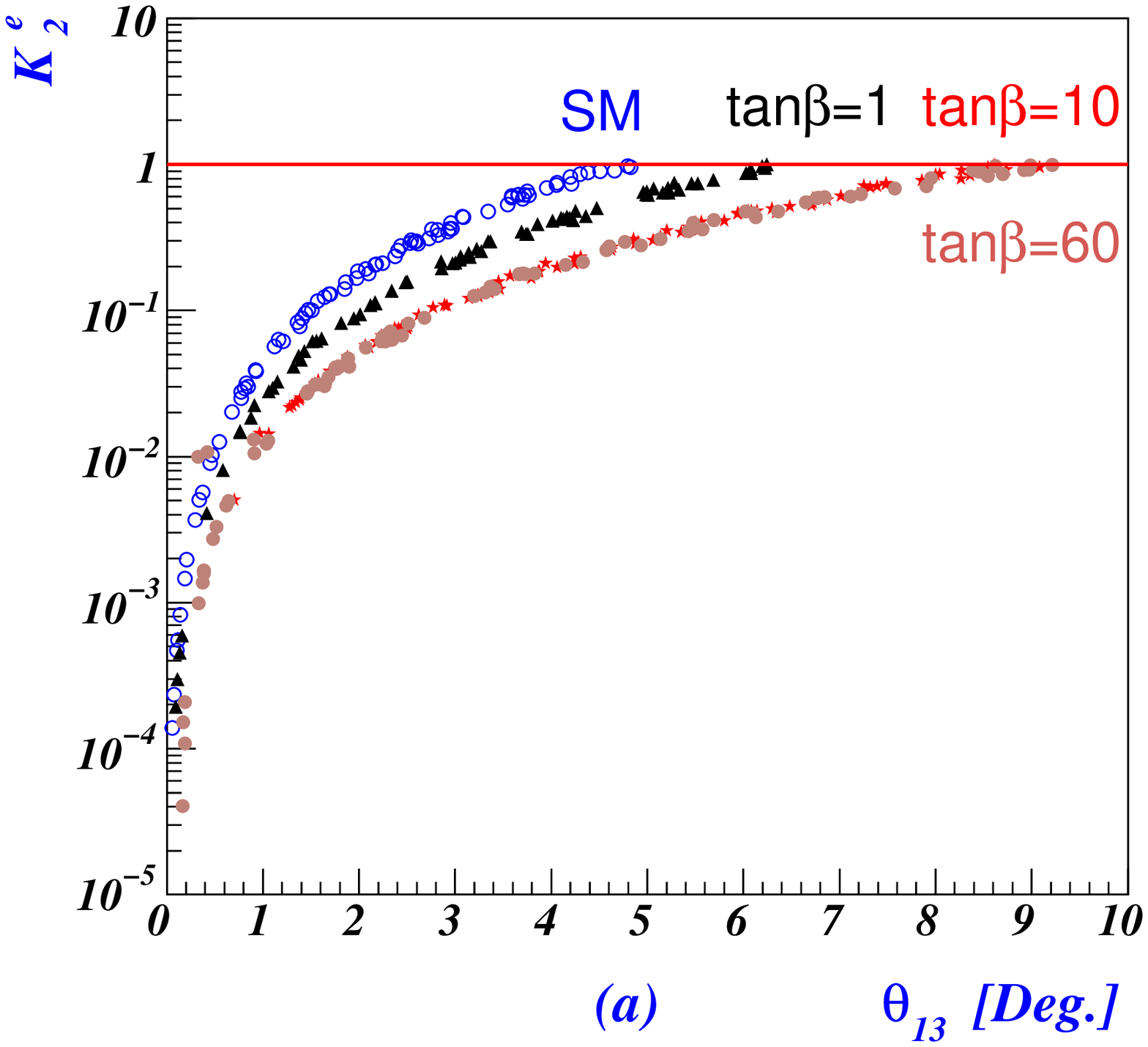,width=6.5cm,angle=0}
\end{minipage}
\hspace*{1.0cm}
\begin{minipage}[t]{6.0cm}
\epsfig{figure=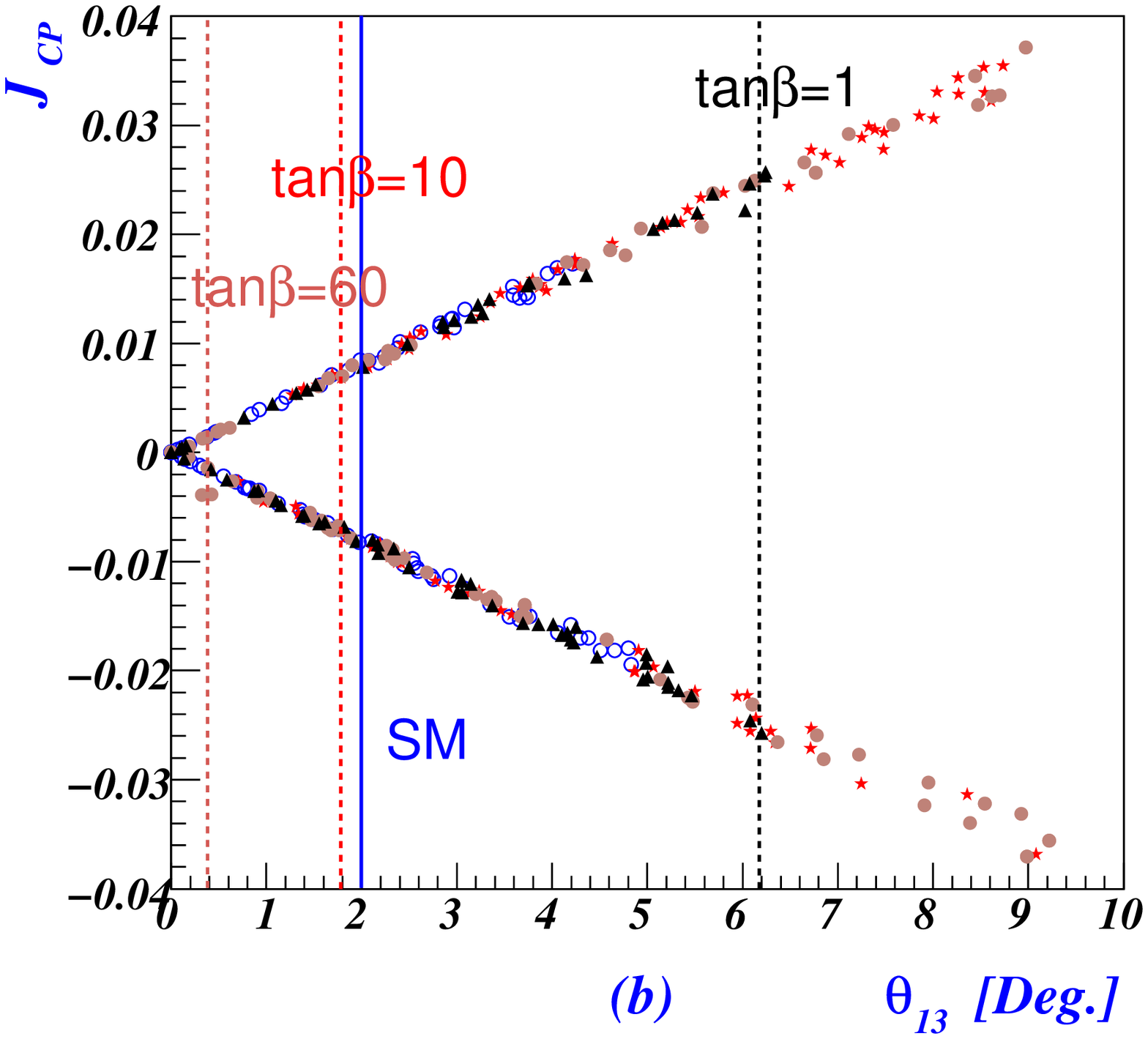,width=6.5cm,angle=0}
\end{minipage}
\caption{\label{Fig2} Plots of wash-out parameter $K^{e}_{2}$ (a) and $J_{\rm CP}$
 as a function of $\theta_{13}$ for the SM (circles), the SSM with $\tan\beta=1~ (\mbox{triangles}), 10~(\mbox{stars}),
  60~(\mbox{spots})$. The vertical dotted lines indicate
the values of $\theta_{13}$ corresponding to the best-fit value of $\eta_B$ for $\tan\beta=1, 10$ and 60, respectively, whereas the
vertical solid line corresponds to the case of the SM.}
\end{figure}

In Fig.~\ref{Fig1}, we show the scatter plots for the model parameters in the SM (upper panel) and the SSM ($\tan\beta=10$) (lower panel)
constrained by the experimental data given in Eq.~(\ref{exp bound}): the figures in the left-hand side exhibit how the
parameters $\kappa~({\rm triangles})$, $\omega~(\rm circles)$, $\lambda~({\rm asters})$ and $\chi({\rm crosses})$ are correlated
with the angle $\alpha$ defined in Eq.~(\ref{tan2alpha}), respectively and the figures in the right-hand side represent how
the difference between the phases $\phi_1$ and $\phi_2$ in Eq.~(\ref{input1}), $|\Delta\phi|~(=|\phi_{2}-\phi_{1}|)$, depends on
the angle $\alpha$ .

In Fig.~\ref{Fig2}, we present how the wash-out factor $K_2^e$ (Fig. 2-a) and CP violating observable $J_{\rm CP}$ (Fig. 2-b) are correlated with the mixing angle $\theta_{13}$. The circles, triangles, stars and spots in Fig. 2 correspond to the
cases of the SM, and  the SSM with $\tan\beta=1$, $\tan\beta=10$ and $
\tan\beta=60$, respectively. {}From Fig. 2-(a), we see that in the case $K_2^e<1$ (weak wash-out regime) the mixing angle $\theta_{13}$ is limited.
Even in the case of the SSM with $\tan\beta =60$, the upper limit of $\theta_{13}$ is about 9 degree. It is remarkable from Fig. 2-(a) that  $\eta^{\rm f}_{B}$ is expected to be enhanced through $K^{e}_{2}$ as $\theta_{13}$ gets lower. Thus, this case with $K_2^e<1$ favors low values of $\theta_{13}$.
The vertical dotted lines in Fig. 2-(b)
indicate the values of $\theta_{13}$ corresponding to best-fit value
of baryon asymmetry of our universe, ${\it i.e.} ~\eta_B^{\rm exp-BT}=6.2\times 10^{-10}$,  when $\tan\beta=1, 10$ and 60, respectively, whereas the
vertical solid line corresponds to the case of the SM. As can be seen in Fig.~\ref{Fig2}-(b),  the prediction of $J_{\rm CP}$ for the
SSM with a low $\tan\beta$  can be measurable in the near future
from long baseline neutrino oscillations.


\begin{figure}[t]
\hspace*{-2cm}
\begin{minipage}[t]{6.0cm}
\epsfig{figure=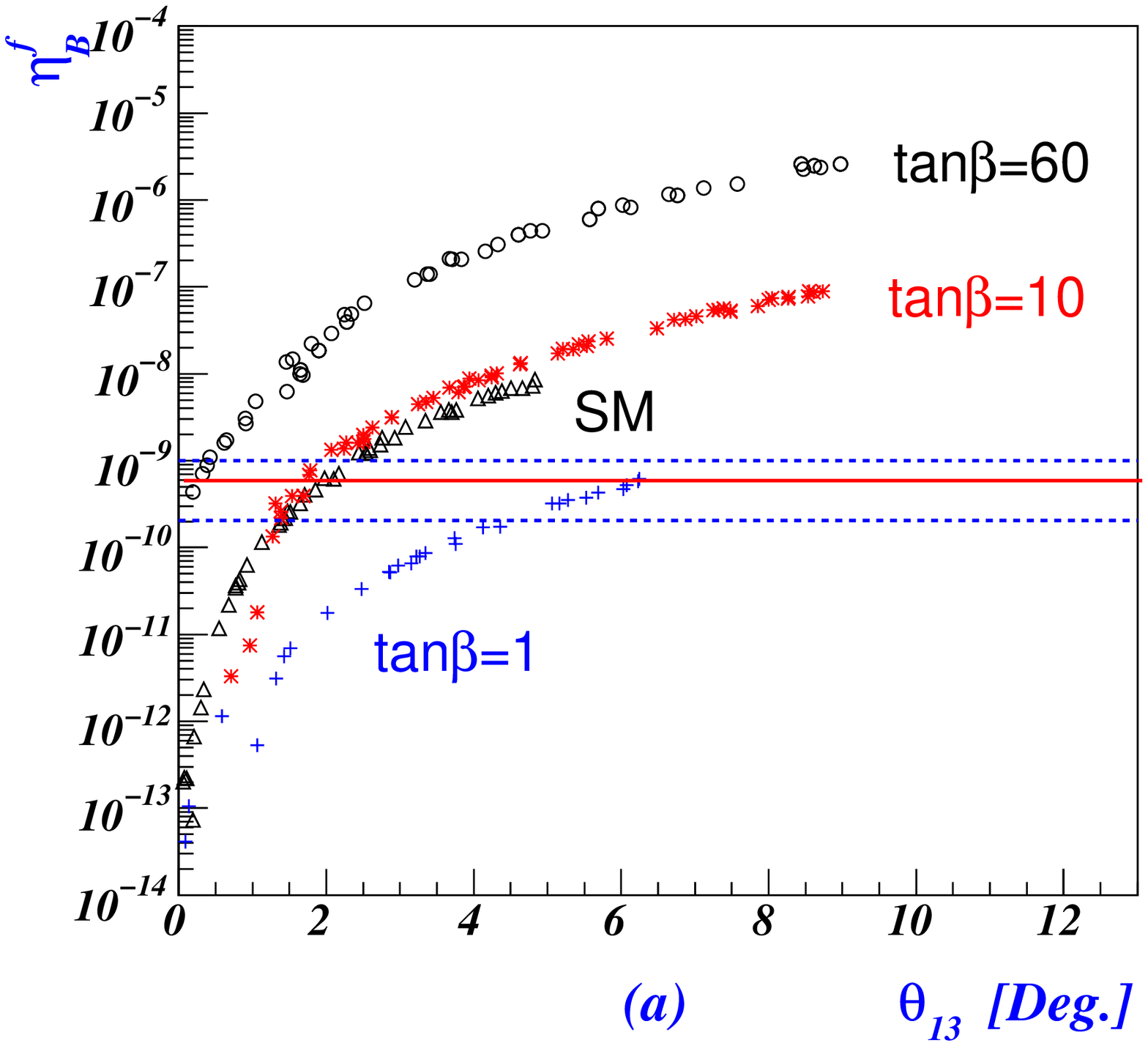,width=6.5cm,angle=0}
\end{minipage}
\hspace*{1.0cm}
\begin{minipage}[t]{6.0cm}
\epsfig{figure=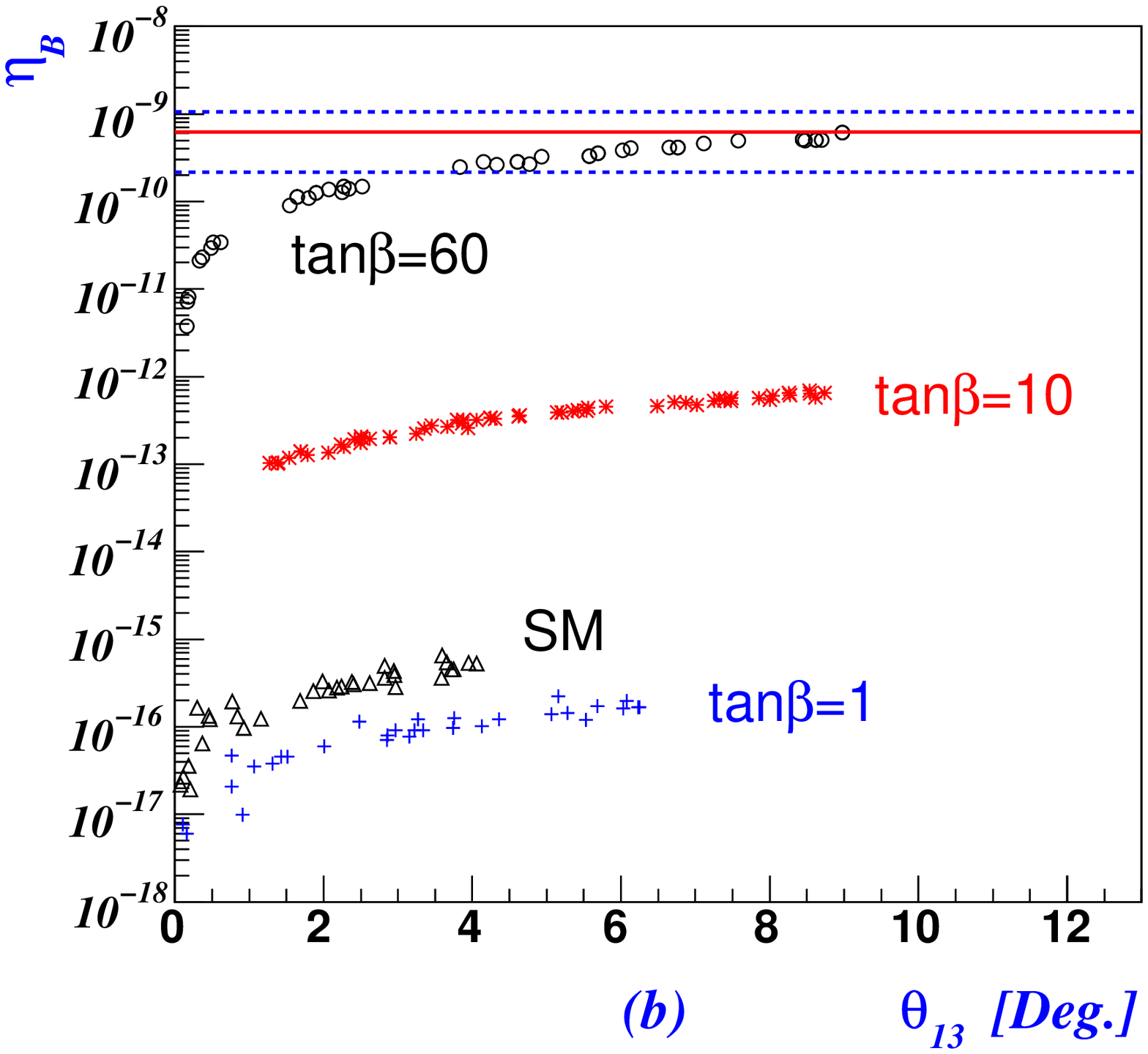,width=6.5cm,angle=0}
\end{minipage}
\caption{ \label{Fig3}  Predictions for the baryon asymmetries (a) $\eta_B^f$ and (b) $\eta_B$  as a function of $\theta_{13}$  for the
same parameter space as in Fig. \ref{Fig1}. The horizontal
dotted lines and solid line in Fig.~\ref{Fig3} (also in Figs.~\ref{Fig4} and Fig.~\ref{Fig6}) correspond to the current
bounds on $\eta_B$, measured from
the current astrophysical observations \cite{cmb}.
}
\end{figure}
\begin{figure}[t]
\hspace*{-2cm}
\begin{minipage}[t]{6.0cm}
\epsfig{figure=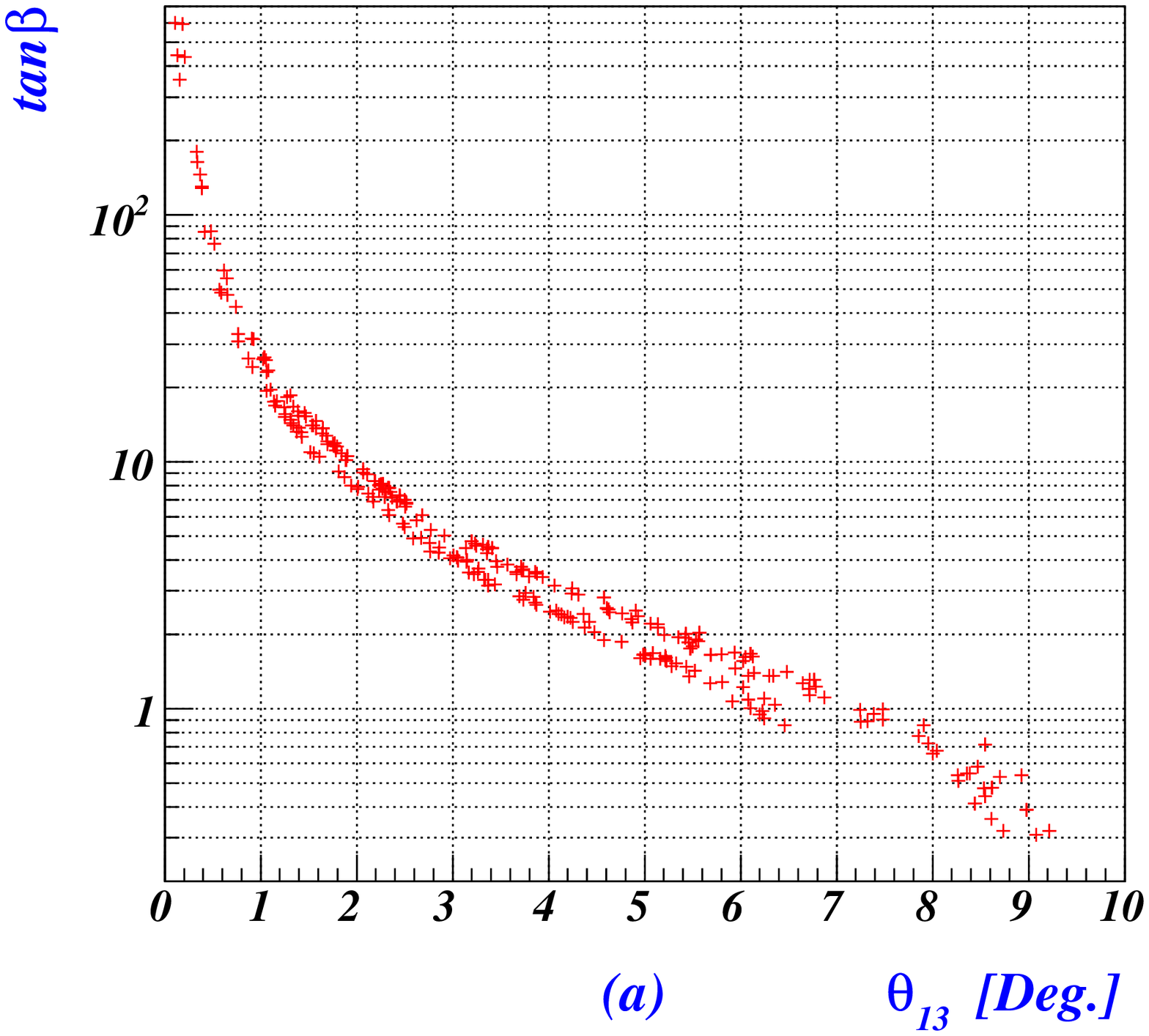,width=6.5cm,angle=0}
\end{minipage}
\hspace*{1.0cm}
\begin{minipage}[t]{6.0cm}
\epsfig{figure=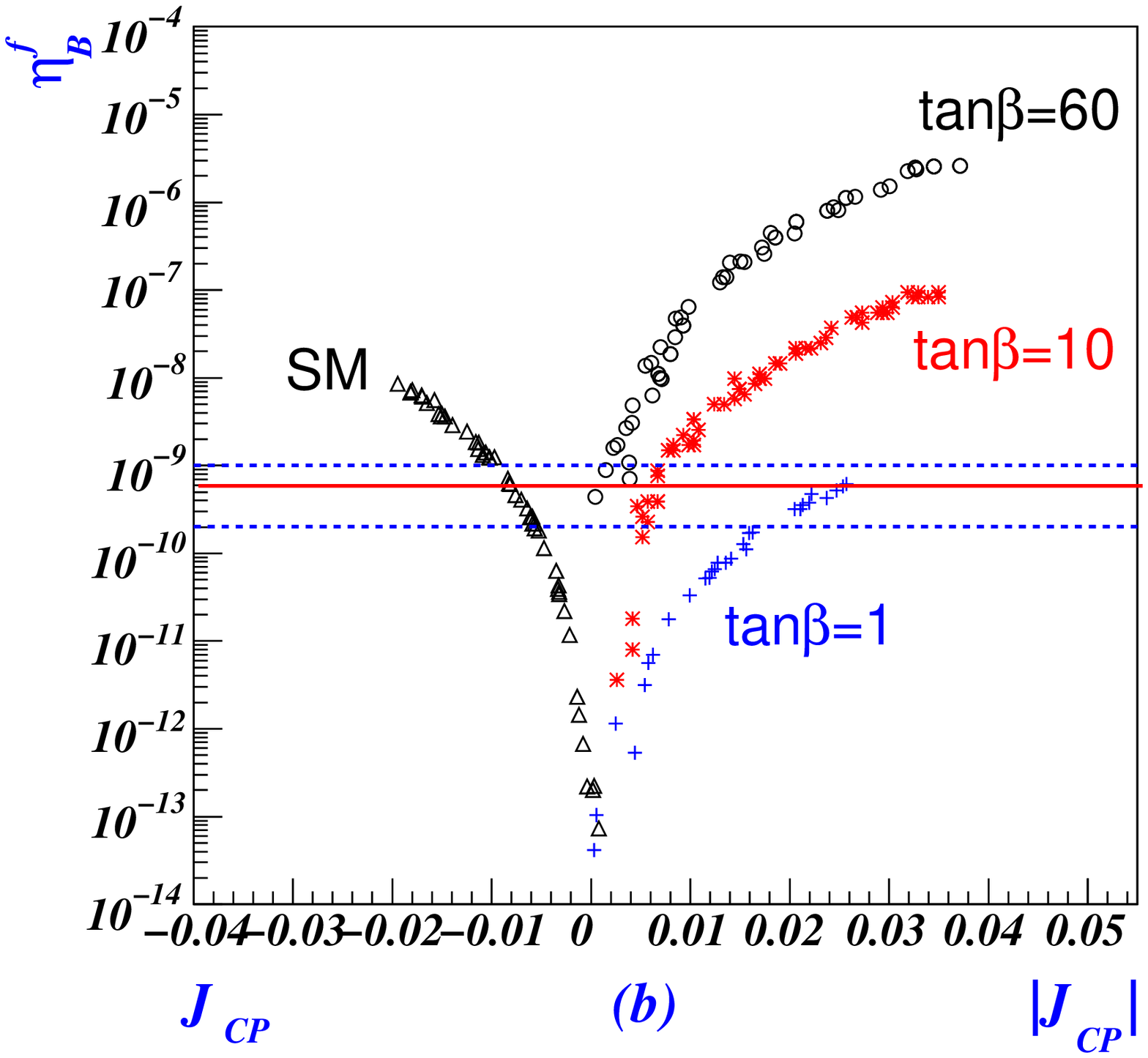,width=6.5cm,angle=0}
\end{minipage}
\caption{\label{Fig4}  (a) Predictions of
$\tan\beta$ in terms of $\theta_{13}$ for flavored leptogenesis.
(b) A link between  $\eta^{\rm f}_B$
and $J_{\rm CP}$ and $|J_{\rm CP}|$ for the SM and the SSM, respectively. }
\end{figure}

In Fig.~\ref{Fig3}, we present the predictions of (a) $\eta^{\rm f}_{B}$  and (b) $\eta_{B}$  for the SM (triangle), the SSM with
$\tan\beta=1$ (daggers), 10 (asters) and 60 (circles) as a function of the mixing angle $\theta_{13}$.
In the case of flavor dependent leptogenesis, the allowed values of $\eta^{\rm f}_B$  prefers very low values of $\theta_{13}$.
The horizontal
dotted lines and solid line in Fig.~\ref{Fig3} (also in Figs.~\ref{Fig4} and Fig.~\ref{Fig6}) correspond to the current bounds on $\eta_B$.
We see from Fig.~\ref{Fig3} that
the prediction of $\eta^{\rm f}_{B}$ in the SSM for a fixed $\theta_{13}$ is increased as $\tan\beta$ increases.
 As explained with Eq.~(\ref{ratioBaryonSSM}), we see from Fig.~\ref{Fig3} that $\eta_B$ is suppressed by factor $4\sim 8$
 compared with $\eta^{\rm f}_B$ because of the different wash-out factors presented in Eq.~(\ref{K-eM}).
 We also see from Fig.~\ref{Fig3} that a successful baryogenesis can be achieved via flavor dependent leptogenesis
 in both the SM and the SSM, whereas a successful baryogenesis via flavor independent leptogenesis is possible only for the SSM
 with large $\tan\beta~(\sim 60)$. For this case,
 we can obtain lower limit of $\tan\beta$ from the current observation for $\eta_{B}$ based
 on Eqs.~(\ref{cmatrix},\ref{h},\ref{deltaM},\ref{RGparameter}) as follows;
  \begin{eqnarray}
   \tan\beta\gtrsim\Big(\frac{2\times10^{3}}{t}\Big[\frac{\tilde{\delta}h_{1}}{\delta^{R}_{\mu\tau}\delta^{I}_{\mu\tau}}\Big]\Big)^{1/4}~.
  \end{eqnarray}

Imposing the best-fit value of observed baryon asymmetry, $\theta_{13}$ is determined to be $2^{\circ}$ for the SM and
$6.2^{\circ},~ 1.8^{\circ}$ and $0.4^{\circ}$ for the SSM with $\tan\beta=1, 10$ and $60$, respectively.
For the case $K^{e}_{2}<1$, the dominant contribution to $\eta^{\rm f}_{B}$ comes from the $e$-leptogenesis $\eta^{e}_{B}$, as shown in Eq.~(\ref{baryonasymSSM}).
In the case of the SSM,  the value of $\theta_{13}$ corresponding to $\eta_B^{\rm exp-BT}$  becomes lower as $\tan\beta$ gets higher
because $K^{e}_{2}$ is proportional to $\theta_{13}$. {}From Eqs.~(\ref{FlaLepto1},\ref{baryonasymSSM},\ref{etaBexp}),
we can roughly get the value of $\tan\beta$ corresponding to $\eta_B^{\rm exp-BT}$ as follows:
  \begin{eqnarray}
   \tan\beta\simeq\Big(\frac{0.1}{K^{e}_{2}K_{2}}\Big[\frac{\tilde{\delta}h_{2}}{\delta^{R}_{\mu\tau}\delta^{I}_{\mu\tau}}\Big]-1\Big)^{1/2}~.
   \label{tanbWK}
  \end{eqnarray}
In Fig.~\ref{Fig4}-(a), we present how $\tan\beta$ is correlated with the  mixing angle $\theta_{13}$, which reflects the above
equation (\ref{tanbWK}).
Fig.~\ref{Fig4}-(b) shows how the baryon asymmetry $\eta^{\rm f}_B$ is sensitive to the CP violating observable
$J_{\rm CP}$ given by Eq.~(\ref{CP1}). The triangles, daggers, asters and circles
correspond to the cases of the SM and the SSM with $\tan\beta=1,10,60$, respectively.
The value of $\eta^{\rm f}_B$ corresponding to $\eta^{\rm exp-BT}_B$ for the case of the SM leads to $|J_{\rm CP}|\simeq 0.008$ (and
$\theta_{13}\simeq2^{\circ}$), which can be measured in the future long-baseline neutrino oscillation experiments.
{}From Fig.~\ref{Fig4}-(b) and Fig.~\ref{Fig2}-(b),  we see that the value of $J_{\rm CP}$ corresponding to $\eta^{\rm exp-BT}_B$
for the case of the SSM becomes higher as $\tan\beta$ gets smaller. Thus, if future experiments can measure $J_{\rm CP}$
with a sensitivity better than 0.01, the predictions of $\eta^{\rm f}_B$ in the SSM with low $\tan\beta~(<10)$ could be probed.

\subsection{In case of $K^{\alpha}_{1,2}>1,~(\alpha=e,\mu,\tau)$}

\begin{figure}[t]
\hspace*{-2cm}
\begin{minipage}[t]{6.0cm}
\epsfig{figure=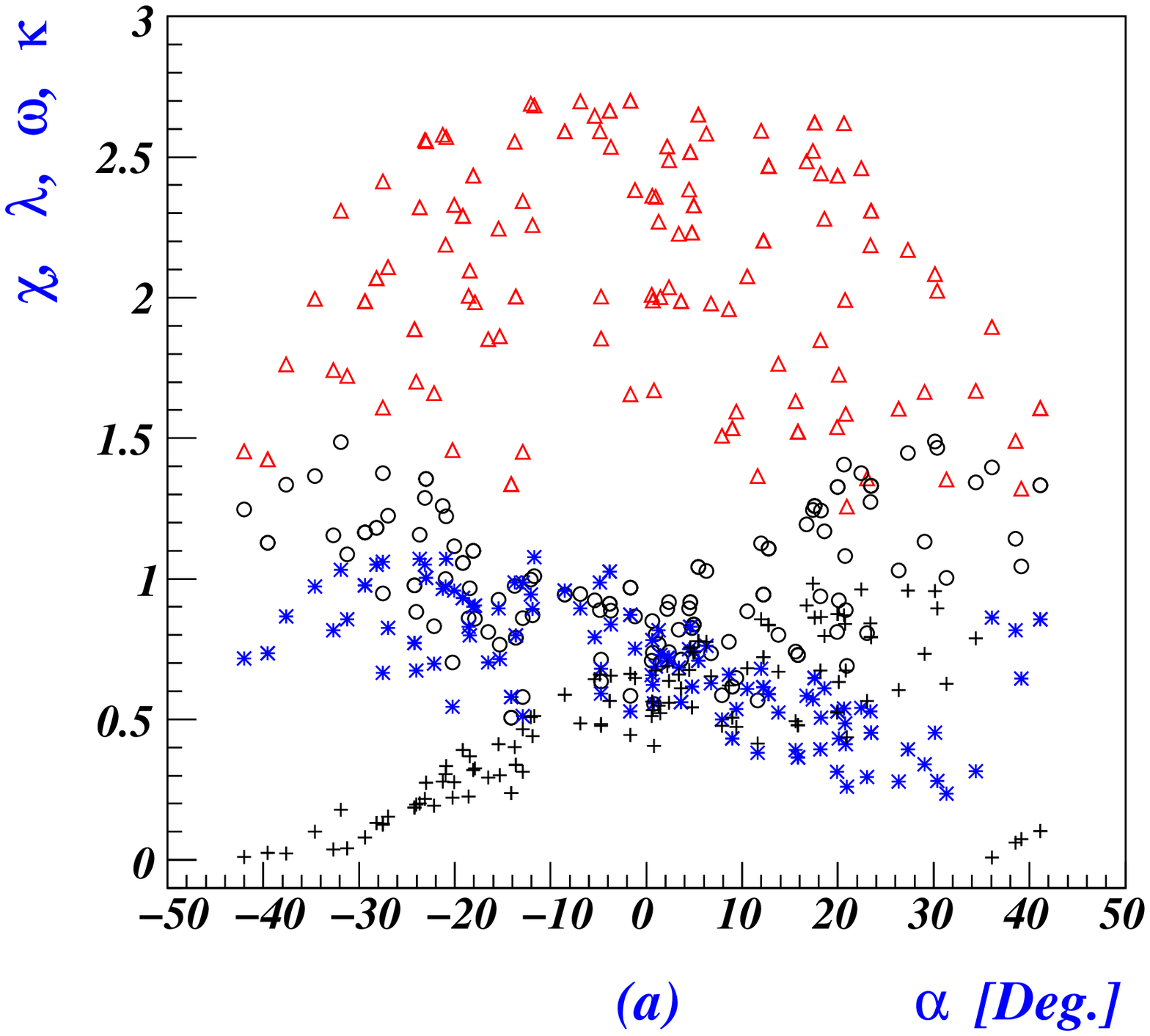,width=6.5cm,angle=0}
\end{minipage}
\hspace*{1.0cm}
\begin{minipage}[t]{6.0cm}
\epsfig{figure=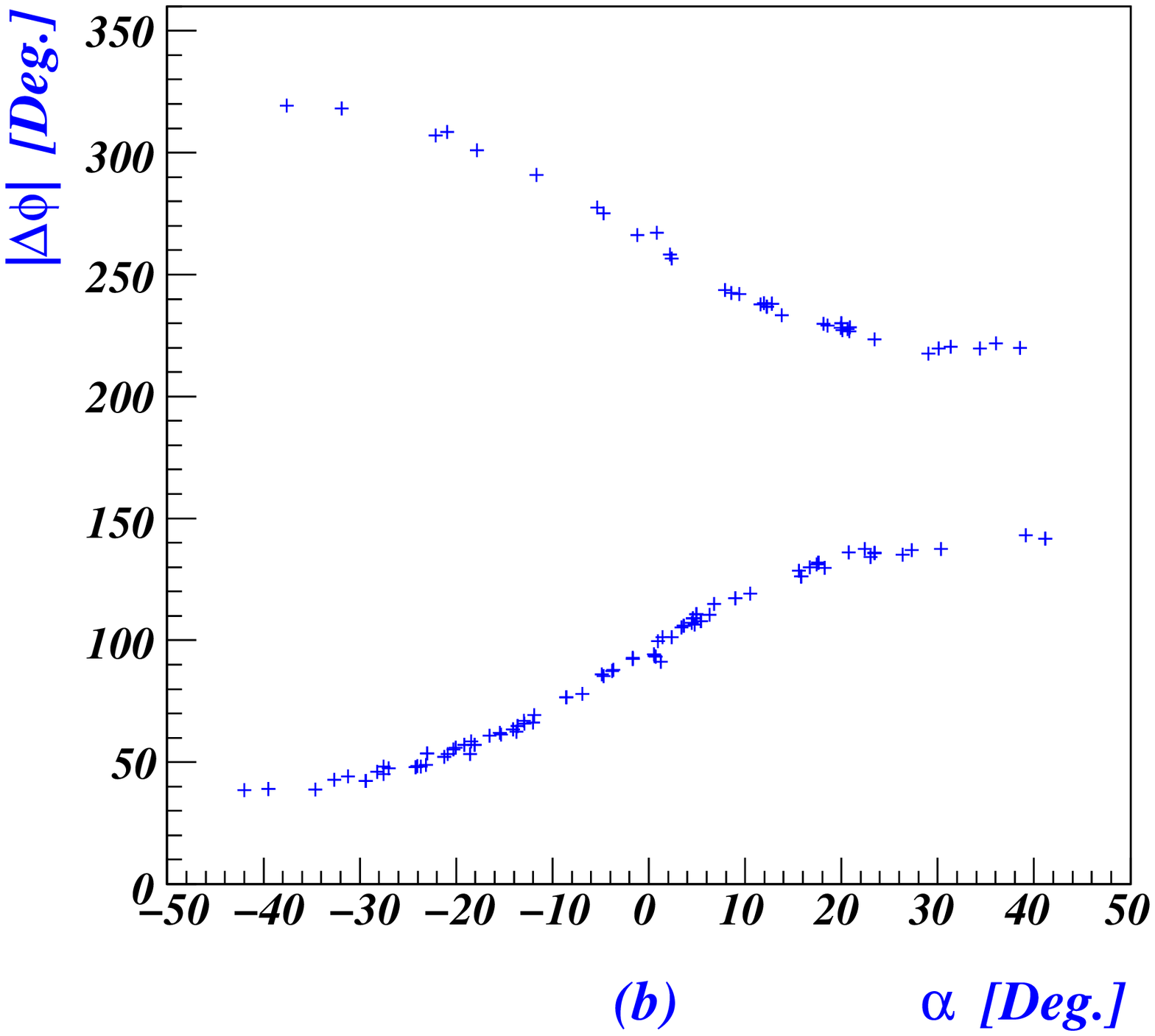,width=6.5cm,angle=0}
\end{minipage}
\caption{\label{Fig5}  The same as Fig.~\ref{Fig1} for
$\tan\beta=10$ except for $K^{\alpha}_{i}>1$.}
\end{figure}
\begin{figure}[t]
\hspace*{-2cm}
\begin{minipage}[t]{6.0cm}
\epsfig{figure=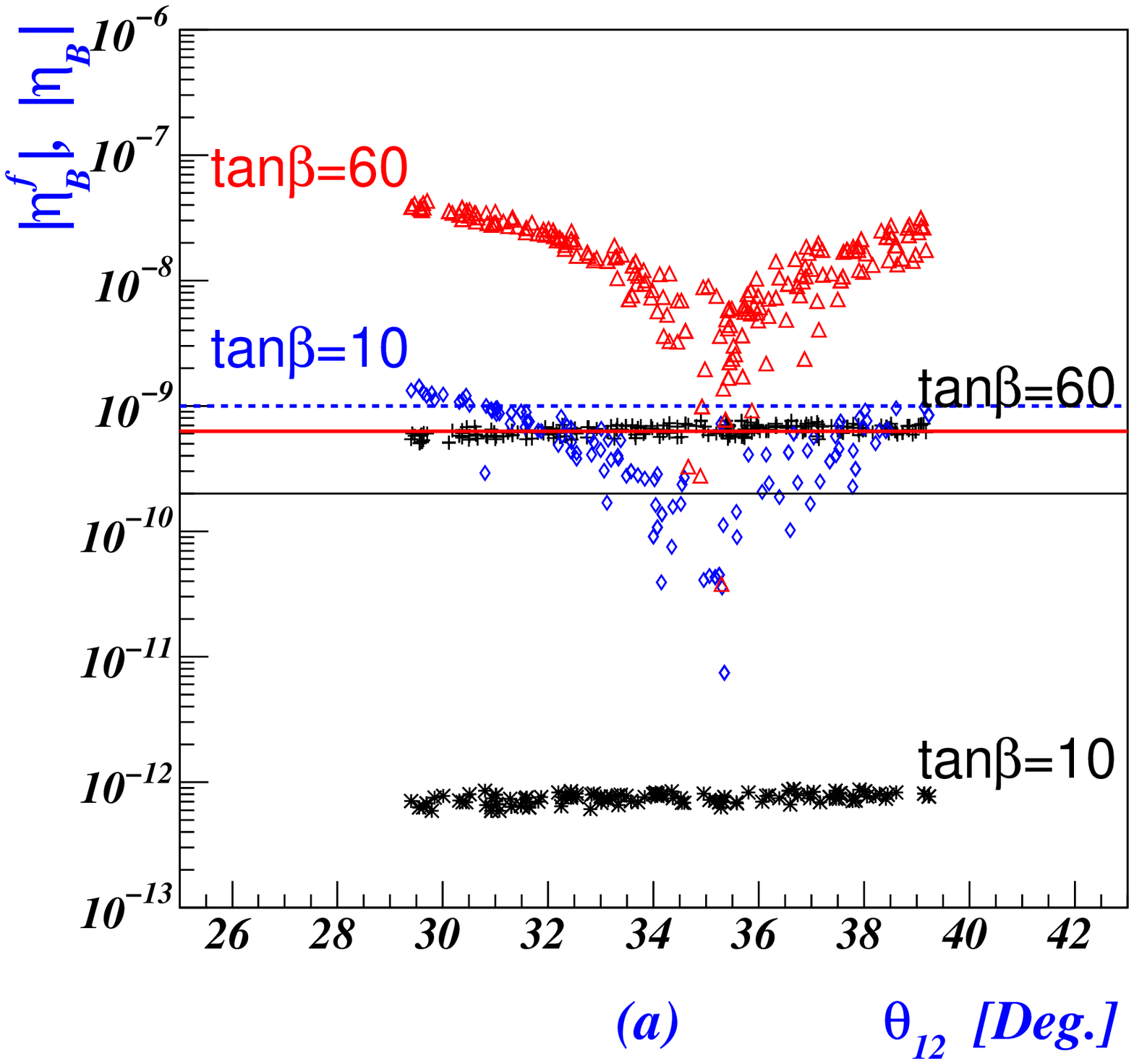,width=6.5cm,angle=0}
\end{minipage}
\hspace*{1.0cm}
\begin{minipage}[t]{6.0cm}
\epsfig{figure=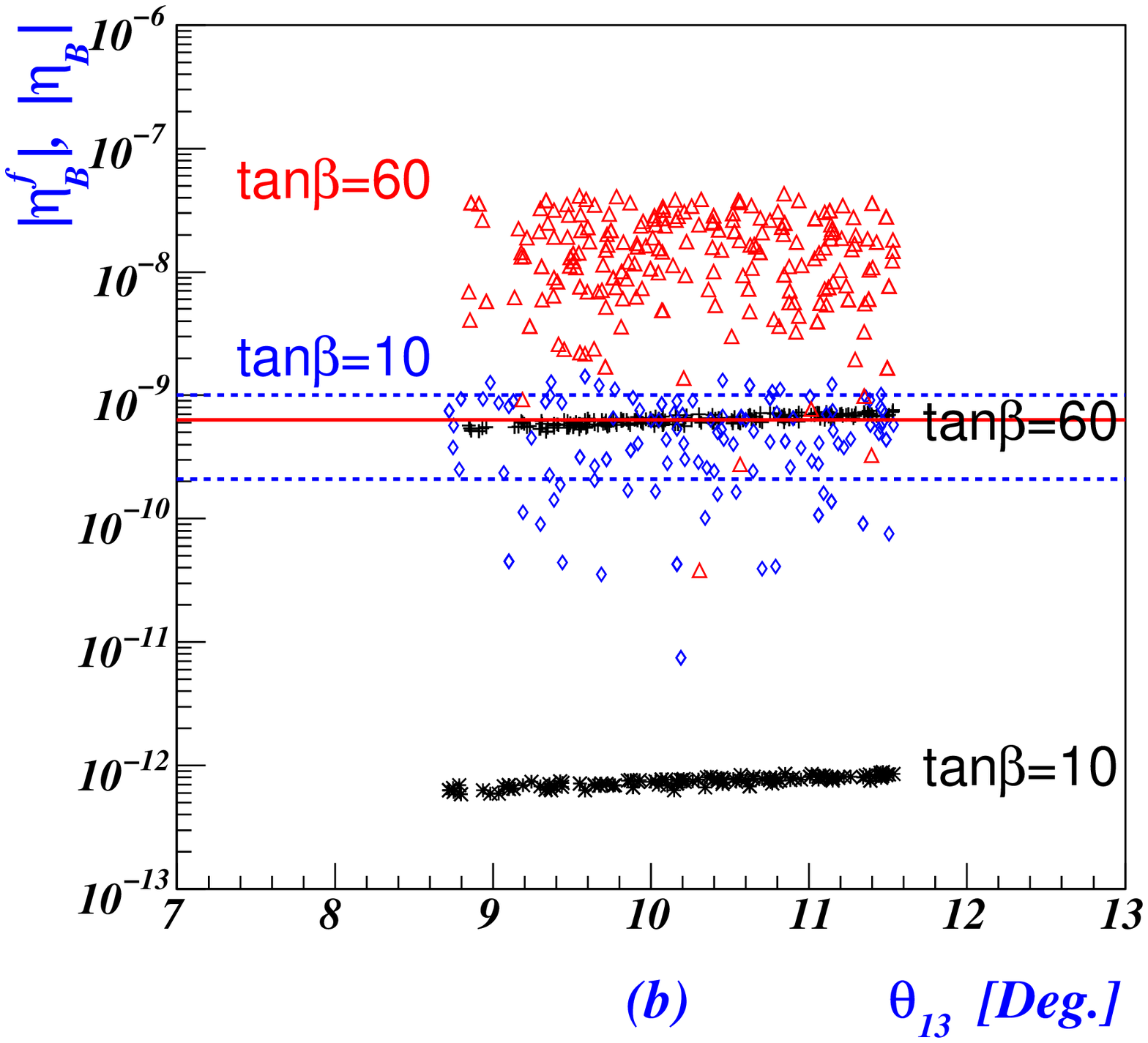,width=6.5cm,angle=0}
\end{minipage}
\caption{\label{Fig6} Predictions for the baryon asymmetry as a function of (a) $\theta_{12}$  and
(b) $\theta_{13}$  for the same parameter space as in Fig.~\ref{Fig5}
in the strong wash-out limit. The triangles
and diamonds (daggers and asters) correspond to flavored (unflavored) leptogenesis
for $\tan\beta=60$ and  $10$, respectively. }
\end{figure}

In the case of strong wash-out for the decay of $N_{1,2}$, $K^{e,\mu,\tau}_{1,2}>1$, it turns out that
baryogenesis in the SM cannot be successfully realized. However, as can
be seen in Eq.~(\ref{baryonasymST}),
flavor dependent leptogenesis in the SSM could successfully be implemented at least
for $\tan\beta\gtrsim7$ as explained below. In Fig.~\ref{Fig5}, we
show the scatter plots for the model parameters in the SSM with
$\tan\beta=10$ constrained by the experimental data given in
Eq.~(\ref{exp bound}): Fig. 5-(a) exhibits how
the parameters $\kappa~({\rm triangles})$, $\omega~(\rm circles)$,
$\lambda~({\rm asters})$ and $\chi~({\rm crosses})$ are correlated
with the angle $\alpha$ defined in Eq.~(\ref{tan2alpha}),
respectively. Fig. 5-(b) represents how
$|\Delta\phi|$ depends on the angle $\alpha$. Fig.~\ref{Fig6}
presents the predictions of baryon asymmetry for the allowed
parameter regions presented in Fig.~\ref{Fig5} as a function of
(a) the solar mixing angle $\theta_{12}$  and (b) the CHOOZ
mixing angle $\theta_{13}$. The results indicate that in the SSM
flavor dependent leptogenesis can successfully work out for $\tan\beta \gtrsim 10$, whereas flavor independent
leptogenesis does so only for the SSM with $\tan\beta \sim 60$.

\begin{figure}[b]
\hspace*{-2cm}
\begin{minipage}[t]{6.0cm}
\epsfig{figure=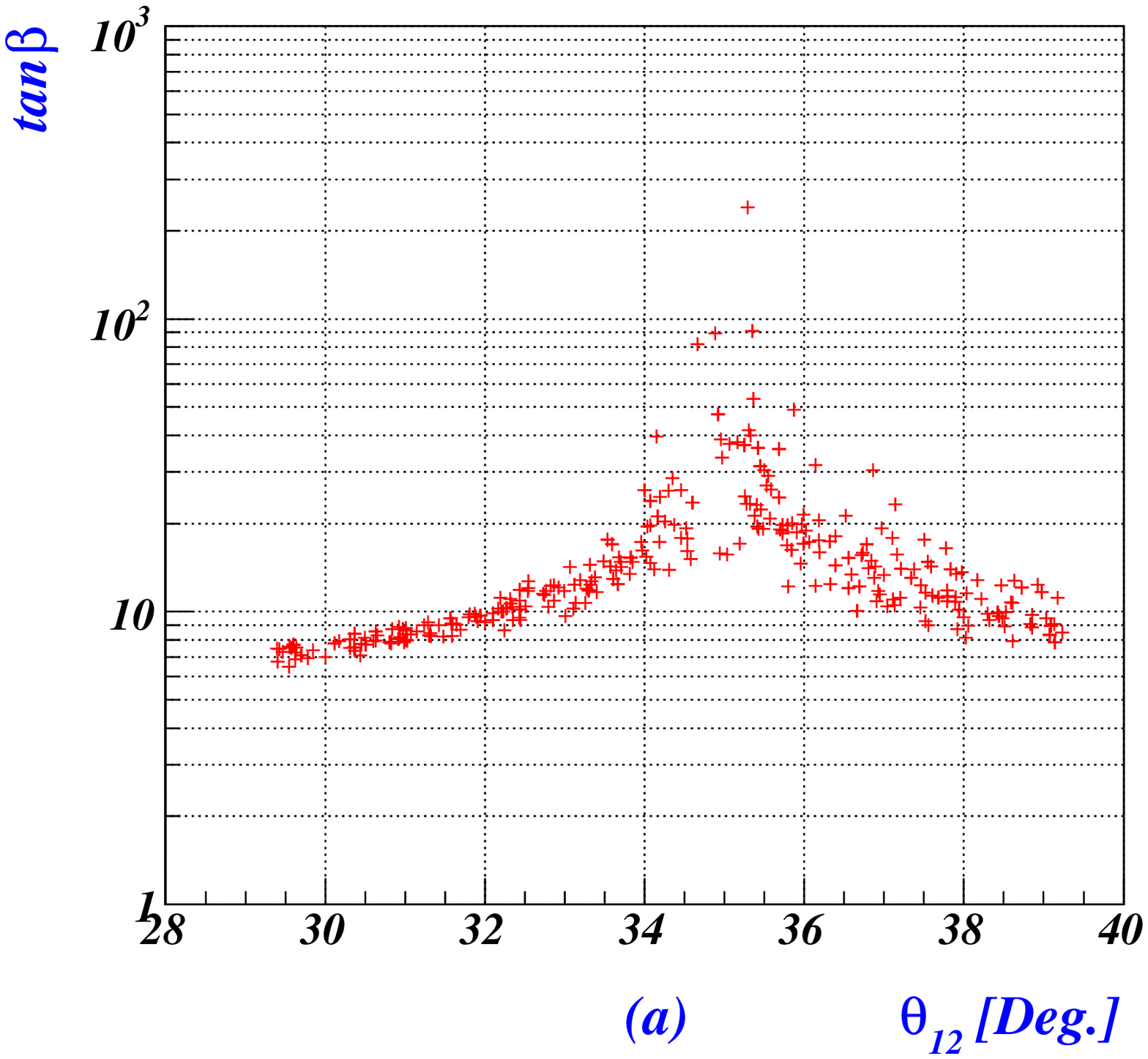,width=6.5cm,angle=0}
\end{minipage}
\hspace*{1.0cm}
\begin{minipage}[t]{6.0cm}
\epsfig{figure=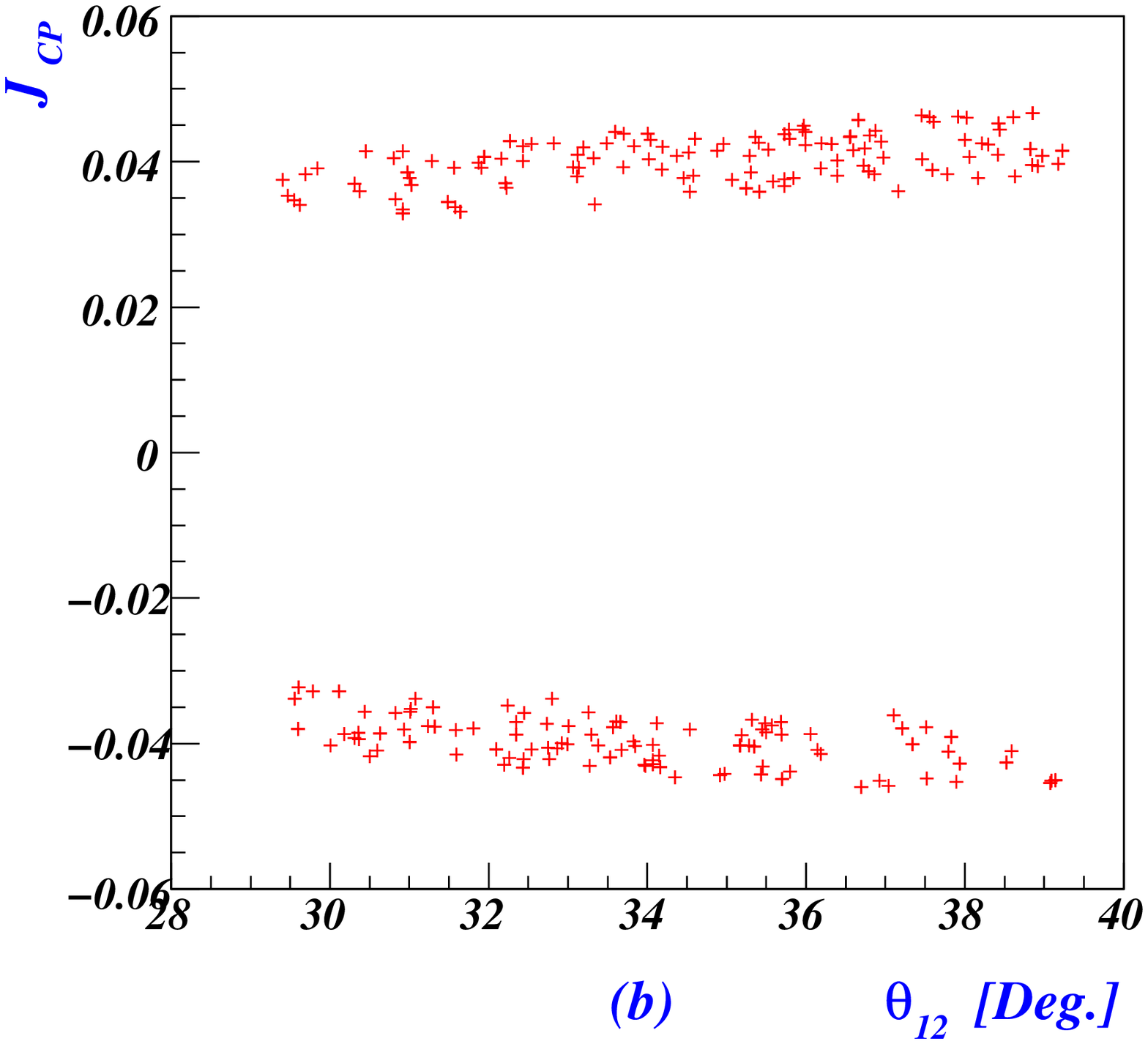,width=6.5cm,angle=0}
\end{minipage}
\caption{\label{Fig7} (a) Plot presenting how  $\tan\beta$ is sensitive to $\theta_{12}$
for flavored leptogenesis. (b) The relationship between $J_{\rm CP}$ and $\theta_{12}$ mentioned in Eq.~(\ref{CP1}).}
\end{figure}
{}From Eqs.~(\ref{FlaLepto1},\ref{etaB},\ref{etaBexp}), we can get the relation,
  \begin{eqnarray}
   \tan\beta\simeq\Big(\frac{\eta^{\rm best-fit}_{B}}{xy^{2}_{\tau\rm SM}}-1\Big)^{1/2}~,
   \label{tanbST}
  \end{eqnarray}
where  $x=\eta^{\rm f}_{B}/y^{2}_{\tau}$.
The relation given by Eq. (\ref{tanbST}) indicates that there exits a correlation between $\tan\beta$ and $\theta_{12}$
for a fixed value of $\eta_B$.
In particular, taking $\eta_B=\eta_B^{\rm exp-BT}$, we plot the correlation between $\tan\beta$ and $\theta_{12}$
in Fig.~\ref{Fig7}-(b).
{}From our numerical estimate, we found that in order for the prediction of $\eta^{\rm f}_B$ in
the case of strong wash-out to be consistent with the right amount of baryon asymmetry given by
 Eq. (\ref{etaBexp}) the value of $\tan\beta$ should be
  \begin{eqnarray}
   \tan\beta\gtrsim7~.
  \end{eqnarray}
Thus, future measurements for $\tan\beta$ and $\theta_{12}$ could serve as an indirect test for the scenario
of baryogenesis considered in this paper.
Fig.~\ref{Fig7}-(b) represents the correlation between $J_{\rm CP}$ and $\theta_{12}$ for a fixed
value of $\eta_B(=\eta^{\rm exp-BT})$.
The results show that the best fit value of $\eta_B$ and the current measurement of $\theta_{12}$
favors $|J_{\rm CP}|\simeq0.04$  which can be measurable in the upcoming long baseline neutrino oscillation experiments.

\section{summary}

As a summary, we have considered an exact $\mu - \tau$  reflection symmetry in neutrino sector
realized at the GUT scale in the context of the seesaw model.
The exact $\mu - \tau$  reflection symmetry has been imposed in the basis where both the charged lepton
mass and heavy Majorana neutrino mass matrices are real and diagonal.
We have assumed that the two lighter heavy Majorana neutrinos are degenerate at the GUT scale.
It has been shown that the RG evolution
from the GUT scale to the seesaw scale gives rise to breaking of the $\mu-\tau$ symmetry and
a tiny splitting between two degenerate heavy Majorana neutrino masses
as well as small variations of the CP phases in $Y_{\nu}$, which are essential to achieve a successful leptogenesis.
Such small RG effects lead to tiny deviations of $\theta_{23}$ from the maximal value  and the CP phase $\delta_{\rm CP}$
from $\frac{\pi}{2}$ imposed at the GUT scale due to $\mu-\tau$ reflection symmetry.
In our scenario, the required amount of the baryon asymmetry $\eta_B$ could be generated via so-called resonant $e$-leptogenesis,
in which the wash-out factor concerned with electron flavor plays a crucial role in reproducing a successful leptogenesis.
 And we have found that the magnitude of the electron leptogenesis is enhanced by $4\sim 8$ orders
due to the flavor effects in the SM, and enhanced further by a factor of $\tan^2 \beta$ for the SSM.

A point deserved to notice is that  CP violation responsible for the generation of baryon asymmetry of our universe comes
from the breaking of both the two degenerate heavy Majorana neutrinos and the CP phases $\phi_{1,2}$ in $Y_{\nu}$ by RG evolutions, which
corresponds to the tiny breakdown of $\mu-\tau$ reflection symmetry, and such small RG effects cause leptogenesis
to directly be linked with the CP violation measurable through neutrino oscillation as well as neutrino
mixing angles $\theta_{12}$ and $\theta_{13}$. We expect that in addition to the reactor and long baseline neutrino experiments
for precise measurements of neutrino mixing angles and CP violation, the measurements for the supersymmetric parameter
$\tan\beta$ at future collider experiments would serve as an indirect test of our scenario of  baryogenesis
based on  the $\mu-\tau$ reflection symmetry.


\acknowledgments{
\noindent YHA was supported by Academia Sinica in Taiwan. TPN was supported  by the Korea Research Foundation
Grant funded by the Korean Government (MOEHRD) No.
KRF-2005-070-C00030. CSK was supported  by  CHEP-SRC Program.
SKK was supported
 by KRF Grant funded by the Korean Government (MOEHRD) No. KRF-2006-003-C00069.
 \\
}


\newpage

\appendix

\section{Relevant Renormalization Group Equations}


 A non-zero leptonic asymmetry can be generated
if and only if the CP odd invariants $\tilde{J}_{CP}={\rm Im Tr}
[H\textbf{M}^{\dag}_{R}\textbf{M}_{R}\textbf{M}^{\dag}_{R}H^{T}\textbf{M}_{R}]$ does not vanish
\cite{Pilaftsis:1997jf, Branco:2001pq}. Since $\tilde{J}_{CP}$ can
be expressed in the form
 \begin{eqnarray}
  \tilde{J}_{\rm CP}=2\sum_{i<j}\Big\{M_{i}M_{j}(M^{2}_{j}-M^{2}_{i}){\rm Im}[H_{ij}]{\rm Re}[H_{ij}]\Big\},~~~H\equiv
  \textbf{Y}_{\nu}\textbf{Y}^{\dag}_{\nu}~,
  \label{CP4}
 \end{eqnarray}
which is relevant for leptogenesis  \cite{Pilaftsis:1997jf}, a non-vanishing leptonic asymmetry requires not only $M_{i}\neq M_{j}$
but also ${\rm Im}[H_{ij}]{\rm Re}[H_{ij}]\neq0, (i\neq j=1,2,3)$, at the leptogenesis scale. Even if we start
from exact degeneracy between the two light heavy Majorana
neutrinos ($N_1, N_2$) at a certain high energy scale, it is likely to see that
some splitting in their masses could be induced at a different
scale (seesaw scale) through RG running effects. And the Dirac neutrino Yukawa matrix $\textbf{Y}_{\nu}$ is also modified
by the same RG effect,
which is very important to obtain nonzero ${\rm Im}[H_{ij}]{\rm Re}[H_{ij}]\neq0$, as will be shown later.

The radiative behavior of the heavy Majorana neutrinos mass matrix $\textbf{M}_{R}$ is dictated by the following RG equations \cite{RG1}:
\begin{eqnarray}
   \frac{d\textbf{M}_{R}}{dt} &=&
     [(\textbf{Y}_\nu \textbf{Y}_\nu^\dagger)\textbf{M}_R+\textbf{M}_R(\textbf{Y}_\nu \textbf{Y}_\nu^\dagger)^T]~,~~~\text{SM}\nonumber\\
   \frac{d\textbf{M}_{R}}{dt} &=&
     2[(\textbf{Y}_\nu \textbf{Y}_\nu^\dagger)\textbf{M}_R+\textbf{M}_R(\textbf{Y}_\nu \textbf{Y}_\nu^\dagger)^T]~,~~\text{SSM}
  \label{RG 1}
 \end{eqnarray}
where
\begin{eqnarray}
t=\frac{1}{16\pi^{2}}\ln(Q/M_{\rm GUT})
\label{scalet}
\end{eqnarray}
 with an arbitrary renormalization scale $Q$. The RG equations for the Dirac-Yukawa neutrino matrix can be written as
 \begin{eqnarray}
   \frac{d\textbf{Y}_{\nu}}{dt} &=&
   \textbf{Y}_{\nu}[(T-\frac{3}{4}g^{2}_2-\frac{9}{4}g^{2}_{1})-\frac{3}{2}(\textbf{Y}^{\dag}_{l}\textbf{Y}_{l}-\textbf{Y}^{\dag}_{\nu}\textbf{Y}_{\nu})]~,~~~\text{SM}\nonumber\\
   \frac{d\textbf{Y}_{\nu}}{dt} &=&
   \textbf{Y}_{\nu}[(T-3g^{2}_2-\frac{3}{5}g^{2}_{1})+\textbf{Y}^{\dag}_{l}\textbf{Y}_{l}+3\textbf{Y}^{\dag}_{\nu}\textbf{Y}_{\nu}~,~~~~~~~\text{SSM}
  \label{RG 2}
 \end{eqnarray}
where $T=Tr(3Y^{\dag}_{u}Y_{u}+\textbf{Y}^{\dag}_{\nu}\textbf{Y}_{\nu})$, $Y_{u}$ and
$\textbf{Y}_{l}$ are the Yukawa matrices for up-type quarks and charged
leptons and $g_{2,1}$ are the ${\rm SU(2)}_{L}$ and ${\rm U(1)}_{Y}$ gauge
coupling constants.

Let us first reformulate the RG equations (\ref{RG 1}) in the basis where $\textbf{M}_R$ is diagonal.
Since $\textbf{M}_R$ is symmetric, it can be diagonalized by a unitary matrix $V$,
\begin{equation}
V^T\textbf{M}_RV = {\rm Diag.}(M_1,M_2, M_3).
\label{RG 3}
\end{equation}
As the structure of the mass matrix $\textbf{M}_R$ changes with the evolution of the scale,
the unitary matrix $V$ depends on the scale, too.
The RG evolution of the matrix $V(t)$ can be written as
\begin{equation}
\frac{d}{dt}V=VA,
\end{equation}
where $A$ is an anti-Hermitian matrix $A^\dagger =-A$ due to the
unitary of $V$. Then, differentiating Eq.~(\ref{RG 3}), we obtain
\begin{eqnarray}
\frac{dM_{i} \delta_{ij}}{dt} &=& A^T_{ij}M_j +M_i A_{ij} + \{ V^T[(\textbf{Y}_\nu \textbf{Y}_{\nu}^\dagger)\textbf{M}_R+\textbf{M}_R(\textbf{Y}_\nu \textbf{Y}_\nu ^\dagger)^T] V\}_{ij}~,~~\text{SM}\nonumber\\
\frac{dM_{i} \delta_{ij}}{dt} &=& A^T_{ij}M_j +M_i A_{ij} + 2\{ V^T[(\textbf{Y}_\nu \textbf{Y}_{\nu}^\dagger)\textbf{M}_R+\textbf{M}_R(\textbf{Y}_\nu \textbf{Y}_\nu ^\dagger)^T] V\}_{ij}~,~\text{SSM}.
\label{RG 4}
\end{eqnarray}
Absorbing the unitary transformation into the Dirac-Yukawa coupling $Y_\nu  \equiv  V^T \textbf{Y}_\nu,$
the real diagonal part of Eq.~(\ref{RG 4}) becomes
\begin{eqnarray}
\frac{dM_i}{dt} &=& 2M_i(Y_\nu Y_{\nu}^\dagger)_{ii}~,~~\text{SM}\nonumber\\
\frac{dM_i}{dt} &=& 4M_i(Y_\nu Y_{\nu}^\dagger)_{ii}~,~~\text{SSM}.
\label{RG 5}
\end{eqnarray}
The off diagonal part of Eq.~(\ref{RG 4}) leads to
 \begin{eqnarray}
  A_{ij}&=& \frac{M_j+M_i}{M_j-M_i}{\rm Re}[(Y_\nu Y_\nu^\dagger)_{ij}]
  +i\frac{M_j-M_i}{M_j+M_i}{\rm Im}[(Y_\nu Y_\nu^\dagger)_{ij}]~,~~~~~\text{SM}\nonumber\\
  A_{ij}&=& 2\frac{M_j+M_i}{M_j-M_i}{\rm Re}[(Y_\nu Y_\nu^\dagger)_{ij}]
  +i2\frac{M_j-M_i}{M_j+M_i}{\rm Im}[(Y_\nu Y_\nu^\dagger)_{ij}]~,~~\text{SSM}.
  \label{RG 6}
 \end{eqnarray}
The RG equations for $Y_{\nu}$ in the basis where $\textbf{M}_{R}$ are diagonal is written as
 \begin{eqnarray}
  \frac{dY_{\nu}}{dt} &=&
  Y_{\nu}[(T-\frac{3}{4}g^{2}_2-\frac{9}{4}g^{2}_{1})-\frac{3}{2}(\textbf{Y}^{\dag}_{l}\textbf{Y}_{l}-Y^{\dag}_{\nu}Y_{\nu})]+A^{T}Y_{\nu}~,~~~\text{SM}\nonumber\\
  \frac{dY_{\nu}}{dt} &=&
  Y_{\nu}[(T-3g^{2}_2-\frac{3}{5}g^{2}_{1})+\textbf{Y}^{\dag}_{l}\textbf{Y}_{l}+3Y^{\dag}_{\nu}Y_{\nu})]+A^{T}Y_{\nu}~,~~~~~~\text{SSM}.
  \label{RG 7}
 \end{eqnarray}
 The RG equations for the quantity $H$ relevant for leptogenesis can be written as
 \begin{eqnarray}
  \frac{dH}{dt}&=&
  2\big(T-\frac{3}{4}g_2^2-\frac{9}{4}g_1^2\big)H-3Y_\nu(\textbf{Y}_l^\dagger \textbf{Y}_l)Y_\nu^\dagger+3H^2+A^TH+HA^\ast~,~\text{SM}\nonumber\\
  \frac{dH}{dt}&=&
  2\big(T-3g^{2}_2-\frac{3}{5}g^{2}_{1}\big)H+2Y_\nu(\textbf{Y}_l^\dagger \textbf{Y}_l)Y_\nu^\dagger+6H^2+A^TH+HA^\ast~,~~~\text{SSM}.
  \label{RG 8}
 \end{eqnarray}
 We see from Eq.~(\ref{RG 6}) that the real part of $A_{ij}$ is singular when $M_i = M_j$.
 The singularity in ${\rm Re}[A_{ij}]$ can be eliminated with the help of an appropriate rotation between degenerate heavy Majorana neutrino states.
Such a rotation does not change any physics and it is equivalent to absorb the rotation matrix $R$ into the Dirac-Yukawa neutrino matrix $Y_{\nu}$,
 \begin{eqnarray}
   Y_\nu \longrightarrow \widetilde{Y}_\nu=RY_\nu,
 \end{eqnarray}
 where the matrix $R$, in our case, particularly rotating the 1st and 2rd generations of heavy Majorana
 neutrinos  can be parameterized as
 \begin{eqnarray}
  R(\alpha) =
  \left(\begin{array}{ccc}
  \cos\alpha & \sin\alpha &0 \\
  -\sin\alpha & \cos\alpha &0  \\
  0 & 0 & 1
  \end{array}\right)~.
  \label{R matrix}
 \end{eqnarray}
 Then, the singularity in the real part of $A_{ij}$ is indeed removed when the rotation angle $\alpha$ is
 taken to be satisfied with the condition,
 \begin{eqnarray}
  {\rm Re}[(\widetilde{Y}_\nu\widetilde{Y}_\nu^\dagger)_{ij}]=0~,~~\text{for any pair}~i,j ~\text{corresponding to}~M_{i}=M_{j}~,
  \label{singularity}
 \end{eqnarray}
 for $i,j=1,2$, which leads to
\begin{eqnarray}
\tan2\alpha &=& \frac{2H_{12}}{H_{11}-H_{22}}=\frac{2(\lambda
\chi+2\omega\kappa\cos\Delta \phi_{12})}{\lambda^2+2\omega^2-\chi^2 -2\kappa^2}~.
\label{tan2alpha}
\end{eqnarray}
With $\widetilde{Y}_\nu$,  we construct a
parameter $\widetilde{H}$ as follows;
\begin{eqnarray}
  \widetilde{H} &\equiv& \widetilde{Y}_\nu\widetilde{Y}_\nu^\dagger = RHR^T= \left(\begin{array}{ccc}
  \widetilde{H}_{11}  &  0  & \widetilde{H}_{13}  \\
  0 &  \widetilde{H}_{22} & \widetilde{H}_{23}  \\
  \widetilde{H}_{13}  &  \widetilde{H}_{23}  & H_{33} \\
 \end{array}
 \right),
   \label{H'1}
\end{eqnarray}
where $H=Y_{\nu}Y^{\dagger}_{\nu}$ and the components of
$\widetilde{H}$ are given by
 \begin{eqnarray}
  \widetilde{H}_{11} &=& H_{11}\cos^2\alpha +H_{12}\sin 2\alpha+H_{22}\sin^2\alpha~,~~~~
  \widetilde{H}_{13} = H_{13}\cos\alpha+H_{23}\sin\alpha~,\nonumber\\
  \widetilde{H}_{22} &=& H_{22}\cos^2\alpha -H_{12}\sin 2\alpha +H_{11}\sin^2\alpha~,~~~~~
  \widetilde{H}_{23}=H_{23}\cos\alpha-H_{13}\sin\alpha~.\nonumber
 \end{eqnarray}
It is then obvious from Eq.~(\ref{H'1}) that ${\rm Re}[(\widetilde{Y}_\nu\widetilde{Y}_\nu^\dagger)_{12(21)}]=0$ and thus the singularity in $A_{12(21)}$ does not appear.

Now, let us consider RG effects which may play an important role in successful leptogenesis.
First, we parameterize the mass splitting of the degenerate heavy Majorana neutrinos
in terms of a parameter $\delta_N$ defined by
 \begin{eqnarray}
  \delta_N \equiv 1-\frac{M_{2}}{M_{1}},
  \label{degeneracy0}
 \end{eqnarray}
which is governed by the following RG equations derived from Eq.~(\ref{RG 5}),
 \begin{eqnarray}
  \frac{d\delta_N}{dt} &=& 2(1-\delta_{N})[\widetilde{H}_{11}-\widetilde{H}_{22}] \simeq \frac{4H_{12}}{\sin2\alpha}~,~~~~\text{SM}\nonumber\\
  \frac{d\delta_N}{dt} &=& 4(1-\delta_{N})[\widetilde{H}_{11}-\widetilde{H}_{22}] \simeq \frac{8H_{12}}{\sin2\alpha}~,~~~~\text{SSM}.
  \label{degeneracy1}
 \end{eqnarray}
 The solutions of the RG equations~(\ref{degeneracy1}) are approximately given by
 \begin{eqnarray}
 \delta_N \simeq \left\{
                  \begin{array}{ll}
                    b^{2}_{3}\tilde{\delta}\cdot t, & \hbox{SM}~, \\
                    2b^{2}_{3}\tilde{\delta}\cdot t, & \hbox{SSM}~,
                  \end{array}\right.~~~~~~~~~~~~~~
                  \text{with}~~\tilde{\delta}=\frac{4(\lambda\chi+2\omega\kappa\cos\Delta\phi_{12})}{\sin 2\alpha},
\label{deltaM}
\end{eqnarray}
where we used the parameters defined in Eq.~(\ref{input2}).

 Next, we consider RG running of the Dirac neutrino Yukawa matrix from the GUT scale to the seesaw scale, $Q\simeq M$.
 Since the RG evolution produces non-zero off-diagonal entries in $M_{R}$, it has to be re-diagonalized by a unitary transformation, $M_{R}\rightarrow V^{T}_{R}M_{R}V_{R}={\rm diag}(M_{1}, M_{2}, M_{3})$ , which leads to the rotation, $N_{R}\rightarrow V_{R}N_{R}$ and $Y_{\nu}\rightarrow V_{R}Y_{\nu}$.
Neglecting the corrections proportional to charged-$\mu$ and -$e$ Yukawa coupling, $y_{\mu}$ and $y_{e}$,
we can obtain the RG improved Dirac neutrino Yukawa matrix given for the SM by
 \begin{eqnarray}
  \widetilde{Y}_{\nu}(M) \simeq \left(\begin{array}{ccc}
  y_{11}-\epsilon y_{21}\cdot t &  y_{12}-\epsilon y_{22}\cdot t & y^{\ast}_{12}-(\epsilon y^{\ast}_{22}+\frac{3y^{2}_{\tau}}{2}y^{\ast}_{12})\cdot t  \\
  y_{21}+\epsilon y_{11}\cdot t &  y_{22}+\epsilon y_{12}\cdot t  & y^{\ast}_{22}+(\epsilon y^{\ast}_{12}-\frac{3y^{2}_{\tau}}{2}y^{\ast}_{22})\cdot t   \\
  0  &  y_{32}  & y^{\ast}_{32} \\
  \end{array} \right)~,
  \label{seesaw1}
 \end{eqnarray}
 and for the SSM by
 \begin{eqnarray}
  \tilde{Y}_{\nu}(M) \simeq \left(\begin{array}{ccc}
  y_{11}+\frac{2\epsilon}{3} y_{21}\cdot t &  y_{12}+\frac{2\epsilon}{3} y_{22}\cdot t  & y^{\ast}_{12}+(\frac{2\epsilon}{3} y^{\ast}_{22}+y^{2}_{\tau}y^{\ast}_{12})\cdot t  \\
  y_{21}-\frac{2\epsilon}{3} y_{11}\cdot t &  y_{22}-\frac{2\epsilon}{3} y_{12}\cdot t  & y^{\ast}_{22}-(\frac{2\epsilon}{3} y^{\ast}_{12}-y^{2}_{\tau}y^{\ast}_{22})\cdot t  \\
  0  &  y_{32}  & y^{\ast}_{32} \\
  \end{array} \right)~,
  \label{seesaw11}
 \end{eqnarray}
 where $\cos\alpha=c_{\alpha},~\sin\alpha=s_{\alpha}$, the components of $\widetilde{Y}_{\nu}(M)$ are
 \begin{eqnarray}
  y_{11} &\equiv& b_{3}(\lambda c_{\alpha}+\chi s_{\alpha})~,~~y_{12}\equiv b_{3}(e^{i\phi_{2}}\kappa s_{\alpha}+e^{i\phi_{1}}\omega c_{\alpha})~,\nonumber\\
  y_{21} &\equiv& b_{3}(\chi c_{\alpha}-\lambda s_{\alpha})~,~~y_{22}\equiv b_{3}(e^{i\phi_{2}}\kappa c_{\alpha}-e^{i\phi_{1}}\omega s_{\alpha})~,~~y_{32}\equiv b_{3}~,
 \end{eqnarray}
 and the parameter $\epsilon$ presenting RG corrections is given by
  \begin{eqnarray}
  \epsilon\simeq \frac{3y^{2}_{\tau}\big\{\frac{\kappa^{2}-\omega^{2}}{2} \sin2\alpha+
  \kappa\omega\cos2\alpha\cos\Delta\phi_{12}\big\}}{\tilde{\delta}}~.
  \label{RGparameter}
 \end{eqnarray}
 Note here that the third low of $\widetilde{Y}_{\nu}(M)$ is not changed because $d(\widetilde{Y}_{\nu})_{3k}/dt$ does not depend
 on $A_{12}$. As will be shown later, the parameter $\epsilon$ is proportional to $y^{2}_{\tau\rm SM}$ in the SM and $y^{2}_{\tau\rm SM}(1+\tan^{2}\beta)$ in the SSM and it plays important roles in both leptogenesis and low-energy CP violation.
At the seesaw scale, the combination of the Yukawa-Dirac coupling matrices and the mass matrix of right-handed heavy Majorana neutrino, respectively, can be written as
 \begin{eqnarray}
 && \widetilde{H}(M) \equiv \widetilde{Y}_{\nu}(M)\widetilde{Y}^{\dag}_{\nu}(M)\simeq\left(\begin{array}{ccc}
  \widetilde{H}_{11} &  \widetilde{H}_{12} & \widetilde{H}_{13}  \\
  \widetilde{H}_{21} &  \widetilde{H}_{22} & \widetilde{H}_{23}  \\
  \widetilde{H}_{13} &  \widetilde{H}_{23} & H_{33} \\ \end{array}\right)~,\nonumber\\
 && M_{R} \simeq {\rm diag}(M_{1},M_{2},M_{3})~,~~~\text{with}~M_{2}\lesssim M_{1}\ll M_{3}~.
  \label{seesaw2}
 \end{eqnarray}
 Here the (1,2) and (2,1)-components of $\widetilde{H}(M)$ are the quantities radiatively generated by RG running.
Considering the structure of $\widetilde{H}$ in Eq.~(\ref{H'1}),
up to non-zero leading contributions in the right side of Eq.~(\ref{RG 8}), the RG equations of $\widetilde{H}_{12}$ are given for the SM by
 \begin{eqnarray}
   \frac{d {\rm Re}[\widetilde{H}_{12}]}{dt}&\simeq &
   -3y^{2}_{\tau}{\rm Re}[(\widetilde{Y}_{\nu1\tau}\widetilde{Y}^\ast_{\nu2\tau})]+{\rm Re}[A_{21}](\widetilde{H}_{22}-\widetilde{H}_{11})~,\nonumber\\
   \frac{d {\rm Im}[\widetilde{H}_{12}]}{dt}&\simeq&
   -3y^{2}_{\tau}{\rm Im}[(\widetilde{Y}_{\nu1\tau}\widetilde{Y}^\ast_{\nu2\tau})]~,
 \end{eqnarray}
and for the SSM by
 \begin{eqnarray}
   \frac{d {\rm Re}[\widetilde{H}_{12}]}{dt}&\simeq &
    2y^{2}_{\tau}{\rm Re}[(\widetilde{Y}_{\nu1\tau}\widetilde{Y}^\ast_{\nu2\tau})]+{\rm Re}[A_{21}](\widetilde{H}_{22}-\widetilde{H}_{11})~,\nonumber\\
   \frac{d {\rm Im}[\widetilde{H}_{12}]}{dt}&\simeq&
    2y^{2}_{\tau}{\rm Im}[(\widetilde{Y}_{\nu1\tau}\widetilde{Y}^\ast_{\nu2\tau})]~.
 \end{eqnarray}
Using Eqs.~(\ref{RG 6},\ref{degeneracy1}), radiatively generated $\widetilde{H}_{12}$ is given approximately in terms of the parameters in Eqs.~(\ref{H1}) and (\ref{H'1}), for the SM, by
  \begin{eqnarray}
   {\rm Re}[\widetilde{H}_{12}] &=& {\rm Re}[\widetilde{H}_{21}]\simeq
    -\frac{3}{2}y_\tau^2b_3^2\Big [\frac{1}{2}(\kappa^2 -\omega^2)\sin2\alpha+\kappa\omega\cos 2\alpha\cos\Delta\phi_{12}\Big ]\cdot t~,\nonumber\\
   {\rm Im}[\widetilde{H}_{12}] &=& -{\rm Im}[\widetilde{H}_{21}]\simeq
    3y_\tau^2b_3^2\omega\kappa\sin\Delta\phi_{12}\cdot t~,
  \label{ImReH'}
  \end{eqnarray}
and for the SSM by
  \begin{eqnarray}
   {\rm Re}[\widetilde{H}_{12}] &=& {\rm Re}[\widetilde{H}_{21}]\simeq
    y_\tau^2b_3^2\Big [\frac{1}{2}(\kappa^2 -\omega^2)\sin2\alpha+\kappa\omega\cos 2\alpha\cos\Delta\phi_{12}\Big ]\cdot t~,\nonumber\\
   {\rm Im}[\widetilde{H}_{12}] &=& -{\rm Im}[\widetilde{H}_{21}]\simeq
    -2y_\tau^2b_3^2\omega\kappa\sin\Delta\phi_{12}\cdot t~.
  \label{ImReH'M}
  \end{eqnarray}
Note here that radiatively generated quantity ${\rm Im}[\widetilde{H}_{12}]$ is proportional to $\sin\Delta\phi_{12}$ which
is fixed by the angle $\alpha$ in Eq.~(\ref{tan2alpha}) due to the $\mu-\tau$ reflection symmetry,
and related with low energy observables $\theta_{13}$ and $J_{\rm CP}$ in Eqs.~(\ref{theta13},\ref{CP1}).



\newpage



\begin{thebibliography}{99}
\def\plb#1#2#3{Phys.\ Lett.\   B    {\bf #1}, (#3) #2}
\def\npb#1#2#3{Nucl.\ Phys.\       {\bf B#1}, (#3) #2}
\def\prd#1#2#3{Phys.\ Rev.\        {\bf D#1}, (#3) #2}
\def\prl#1#2#3{Phys.\ Rev.\ Lett.\ {\bf #1},  (#3) #2}
\def\mpl#1#2#3{Mod.\ Phys.\ Lett.\ {\bf A#1}, (#3) #2}
\def\rep#1#2#3{Phys.\ Rep.\        {\bf #1},  (#3) #2}
\def\sci#1#2#3{Science             {\bf #1},  (#3) #2}
\def\astro#1#2#3{Astrophys.\ J.\   {\bf #1},  (#3) #2}
\def\epj#1#2#3{Eur.\ Phys.\ J.  {\bf C#1},  (#3) #2}
\def\jhep#1#2#3{JHEP               {\bf #1},  (#3) #2}
\def\jpg#1#2#3{J.\ Phys.\        {\bf G#1},  (#3) #2}
\def\ijmp#1#2#3{Int.\ J.\ Mod.\ Phys.\ {\bf #1},  (#3) #2}
\def\ptp#1#2#3{Prog.\ Theor.\ Phys.\ {\bf #1},  (#3) #2}

\bibitem{atm}
 Y.~Fukuda {\it et al.} [Super-Kamiokande Collaboration], \prl{81}{1562}{1998}.

\bibitem{SK2002}
 S. Fukuda {\it et al.} [Super-Kamiokande Collab.], \prl{86}{5656}{2001}; \plb{539}{179}{2002}.

\bibitem{SNO}
 Q. Ahmad {\it et al.} [SNO Collab.], \prl{87}{071301}{2001}; Q. Ahmad {\it et al.} [SNO Collab.], \prl{89}{011301}{2002}; S.Ahmed {\it et al.} [SNO Collab.], arXiv:nucl-ex/0309004.

\bibitem{chooz}
 M.~Apollonio {\it et al.} [CHOOZ Collaboration], \plb{420}{397}{1998}.

\bibitem{review}
  M.~Fukugita and T.~Yanagida, Phys.\ Lett.\  B {\bf 174}, (1986) 45;  G.~F.~Giudice {\it et al.}, Nucl.\ Phys.\ B {\bf 685} (2004) 89
  [arXiv:hep-ph/0310123]; W.~Buchmuller, P.~Di Bari and M.~Plumacher, Annals Phys.\  {\bf 315}, (2005) 305
  [arXiv:hep-ph/0401240]; A.~Pilaftsis and T.~E.~J.~Underwood, Phys.\ Rev.\  D {\bf 72}, (2005) 113001
  [arXiv:hep-ph/0506107].

\bibitem{seesaw}
 P.~Minkowski, Phys.\ Lett.\  B {\bf 67}, (1977) 421; M. Gell-Mann, P. Ramond and R. Slansky,  {\em Proceedings of the Supergravity Stony Brook Workshop}, New York 1979,  eds. P. Van Nieuwenhuizen and D. Freedman; T. Yanagida,  {\em Proceedinds of the Workshop on Unified Theories and Baryon Number in the Universe},  Tsukuba, Japan 1979, ed.s A. Sawada and A. Sugamoto; R. N. Mohapatra, G. Senjanovic,  \prl{44}{912}{1980}.

\bibitem{Harrison:2002et}
  P.~F.~Harrison and W.~G.~Scott,
  Phys.\ Lett.\ B {\bf 547}, (2002) 219  [arXiv:hep-ph/0210197].

\bibitem{Grimus}
   Walter Grimus, Luis Lavoura, Phys. Lett. B {\bf 579}, (2004) 113.



\bibitem{A4CP}
 K.~S.~Babu, E.~Ma and J.~W.~F.~Valle, Phys.\ Lett.\ B {\bf 552}, (2003) 207 [arXiv:hep-ph/0206292].




\bibitem{yasue}
 I.~Aizawa, T.~Kitabayashi and M.~Yasue, Phys.\ Rev.\ D {\bf 72}, (2005) 055014 [arXiv:hep-ph/0504172]; I.~Aizawa, T.~Kitabayashi and M.~Yasue, Nucl.\ Phys.\ B {\bf 728}, (2005)  220 [arXiv:hep-ph/0507332];
 Z.~z.~Xing, H.~Zhang and S.~Zhou, \ Lett.\ B {\bf 641}, (2006) 189 [arXiv:hep-ph/0607091].


\bibitem{Farzan}
  Yasaman Farzan, Alexei Yu. Smirnov, JHEP {\bf 01}, (2007)  059  [arXiv:hep-ph/0204360].

\bibitem{factory}
  A.~De Rujula, M.~B.~Gavela and P.~Hernandez, Nucl.\ Phys.\ B {\bf 547}, (1999) 21 [arXiv:hep-ph/9811390]; A.~Cervera, A.~Donini, M.~B.~Gavela, J.~J.~Gomez Cadenas, P.~Hernandez, O.~Mena and S.~Rigolin, Nucl.\ Phys.\ B {\bf 579}, (2000) 17 [Erratum-ibid.\  B {\bf 593}, (2001) 731] [arXiv:hep-ph/0002108]; C.~Albright {\it et al.}, arXiv:hep-ex/0008064; K.~Dick, M.~Freund, M.~Lindner and A.~Romanino, Nucl.\ Phys.\  B {\bf 562}, (1999) 29 [arXiv:hep-ph/9903308]; V.~D.~Barger, S.~Geer, R.~Raja and K.~Whisnant, Phys.\ Rev.\ D {\bf 63}, (2001) 033002 [arXiv:hep-ph/0007181]; M.~Freund, P.~Huber and M.~Lindner, Nucl.\ Phys.\  B {\bf 585}, (2000) 105 [arXiv:hep-ph/0004085]; K.~Hagiwara, N.~Okamura and K.~i.~Senda, arXiv:hep-ph/0607255.


\bibitem{Antusch:2005gp}
  S.~Antusch, J.~Kersten, M.~Lindner, M.~Ratz and M.~A.~Schmidt, 
  JHEP {\bf 0503}, (2005)  024  [arXiv:hep-ph/0501272].


\bibitem{RGE}
  P.~H.~Chankowski and Z.~Pluciennik,  Phys.\ Lett.\   B {\bf 316}, (1993) 312 [arXiv:hep-ph/9306333];
  K.~S.~Babu, C.~N.~Leung and J.~T.~Pantaleone,  Phys.\ Lett.\  B {\bf 319}, (1993) 191 [arXiv:hep-ph/9309223];
  M.~Tanimoto,  Phys.\ Lett.\  B {\bf 360}, (1995) 41 [arXiv:hep-ph/9508247];
  N.~Haba and N.~Okamura,  Eur.\ Phys.\ J.\  C {\bf 14}, (2000) 347 [arXiv:hep-ph/9906481];
  P.~H.~Chankowski and S.~Pokorski,  Int.\ J.\ Mod.\ Phys.\  A {\bf 17}, (2002) 575  [arXiv:hep-ph/0110249].


\bibitem{RG1}
  J.~A.~Casas, J.~R.~Espinosa, A.~Ibarra and I.~Navarro,  Nucl.\ Phys.\  B {\bf 573}, (2000) 652 [arXiv:hep-ph/9910420];
  J.~A.~Casas, J.~R.~Espinosa, A.~Ibarra and I.~Navarro,  Nucl.\ Phys.\  B {\bf 569}, (2000) 82 [arXiv:hep-ph/9905381].


\bibitem{Antusch:2002rr}
  S.~Antusch, J.~Kersten, M.~Lindner and M.~Ratz,
  Phys.\ Lett.\  B {\bf 538}, (2002) 87  [arXiv:hep-ph/0203233].

\bibitem{Pilaftsis:1997jf}
  A.~Pilaftsis, 
  Phys.\ Rev.\  D {\bf 56}, (1997) 5431 [arXiv:hep-ph/9707235];
  A.~Pilaftsis and T.~E.~J.~Underwood, 
  Nucl.\ Phys.\  B {\bf 692}, (2004) 303 [arXiv:hep-ph/0309342].


\bibitem{Ahns1}
Y.H. Ahn, S.K. Kang, C.S. Kim and Jake Lee, Phys.\ Rev.\  D {\bf 73}, (2006) 093005 [arXiv:hep-ph/0602160].


\bibitem{Ahn:2006rn}
  Y.~H.~Ahn, C.~S.~Kim, S.~K.~Kang and J.~Lee,
  Phys.\ Rev.\  D {\bf 75}, (2007) 013012  [arXiv:hep-ph/0610007].

\bibitem{SFKing}
 S.~F.~King, JHEP {\bf 0209}, (2002)  011.



\bibitem{Jarlskog} C. Jarlskog, Phys. Rev. Lett. {\bf 55}, (1985) 1039;
D. D. Wu, Phys. Rev. D {\bf 33}, (1986) 860.



\bibitem{lepto2}
 L. Covi, E. Roulet and F. Vissani, Phys. Lett. {\bf B384}, (1996) 169; A.~Pilaftsis, Int.\ J.\ Mod.\ Phys.\ A {\bf 14}, (1999) 1811 [arXiv:hep-ph/9812256].

\bibitem{Flavor}
T. Endoh, T. Morozumi, Z. Xiong, Prog.\ Theor.\ Phys.\ {\bf 111}, (2004) 123
[arXive:hep-ph/0308276];
T. Fujihara, S. Kaneko, S. K. Kang, D. Kimura, T. Morozumi, M. Tanimoto,  Phys.\ Rev.\
D {\bf 72}, (2005) 016006 [arXive:hep-ph/0505076];
  A.~Abada, S.~Davidson, A.~Ibarra, F.~X.~Josse-Michaux, M.~Losada and A.~Riotto,  JHEP {\bf 0609},  (2006) 010 [arXiv:hep-ph/0605281];
  S.~Blanchet and P.~Di Bari,  JCAP {\bf 0703},  (2007) 018 [arXiv:hep-ph/0607330];
  S.~Antusch, S.~F.~King and A.~Riotto,  JCAP {\bf 0611},  (2006) 011  [arXiv:hep-ph/0609038];
  S.~Pascoli, S.~T.~Petcov and A.~Riotto, Phys.\ Rev.\  D {\bf 75},  (2007) 083511 [arXiv:hep-ph/0609125];
  S.~Pascoli, S.~T.~Petcov and A.~Riotto,
  Nucl.\ Phys.\  B {\bf 774}, (2007) 1 [arXiv:hep-ph/0611338];
  G.~C.~Branco, R.~Gonzalez Felipe and F.~R.~Joaquim,  Phys.\ Lett.\  B {\bf 645}, (2007) 432
  [arXiv:hep-ph/0609297];
  G.~C.~Branco, A.~J.~Buras, S.~Jager, S.~Uhlig and A.~Weiler, JHEP {\bf 0709},  (2007) 004
  [arXiv:hep-ph/0609067].



\bibitem{Ahn:2007mj}
  Y.~H.~Ahn, S.~K.~Kang, C.~S.~Kim and J.~Lee,
  Phys.\ Rev.\  D {\bf 77},  (2008) 073009  [arXiv:hep-ph/0711.1001].

\bibitem{Abada}
  A.~Abada, S.~Davidson, F.~X.~Josse-Michaux, M.~Losada and A.~Riotto,  JCAP {\bf 0604},  (2006) 004 [arXiv:hep-ph/0601083];
   S.~Antusch, S.~F.~King and A.~Riotto,  JCAP {\bf 0611}, (2006)  011  [arXiv:hep-ph/0609038].

\bibitem{PU2}
   A.~Pilaftsis,  Phys.\ Rev.\ Lett.\  {\bf 95},  (2005) 081602 [arXiv:hep-ph/0408103];
  A.~Pilaftsis and T.~E.~J.~Underwood,  Phys.\ Rev.\  D {\bf 72},  (2005) 113001 [arXiv:hep-ph/0506107].


\bibitem{Maltoni:2004ei}
  M.~Maltoni, T.~Schwetz, M.~A.~Tortola and J.~W.~F.~Valle,  New J.\ Phys.\  {\bf 6}, (2004) 122  [arXiv:hep-ph/0405172].


\bibitem{cmb} WMAP Collaboration, D.N. Spergel {\it et al.},  Astrophys. J. Suppl. {\bf 148},  (2003) 175;
 M. Tegmark {\it et al.}, Phys. Rev. D {\bf 69},  (2004) 103501; C.~L.~Bennett {\it et al.}, Astrophys.\ J.\ Suppl.\  {\bf 148}, (2003) 1 [arXiv:astro-ph/0302207].


\bibitem{Branco:2001pq}
  G.~C.~Branco  {\it et al.}, Nucl.\ Phys.\ B {\bf 617}, (2001) 475 [arXiv:hep-ph/0107164]; S.~Davidson, J.~Garayoa, F.~Palorini and N.~Rius,
  Phys.\ Rev.\ Lett.\  {\bf 99}, (2007)  161801 [arXiv:hep-ph/0705.1503].



\end{thebibliography}
\end{document}